\documentclass[
 aps,prl,showpacs,superscriptaddress,numerical,twocolumn,amsmath,amssymb,floatfix,
 reprint%
]{revtex4-1}
\usepackage{xcolor}
\usepackage[
	pdffitwindow=true,
	colorlinks=true,
	frenchlinks=false,
        linkcolor=blue,
	anchorcolor=blue,
        citecolor=blue,
        filecolor=blue,
        urlcolor=blue,
        bookmarks=true,
        bookmarksopen=true,
	bookmarksnumbered=true, 
        bookmarksopenlevel=1,
        plainpages=false,
	pdfpagelayout=OnePage,
        pdfpagelabels=true,
	breaklinks
]{hyperref}
\usepackage[per-mode=symbol,separate-uncertainty]{siunitx}
\usepackage{graphicx}
\usepackage{dcolumn}
\usepackage{bm}
\usepackage[english]{babel}
\usepackage{cases}

\begin{document}
\preprint{AIP/123-QED}

\title[]{Photon Statistics of Propagating Thermal Microwaves}

\author{J.~Goetz}
\email[]{jan.goetz@wmi.badw.de}
\affiliation{Walther-Mei{\ss}ner-Institut, Bayerische Akademie der Wissenschaften, 85748 Garching, Germany }
\affiliation{Physik-Department, Technische Universit\"{a}t M\"{u}nchen, 85748 Garching, Germany}
\author{S.~Pogorzalek}
\affiliation{Walther-Mei{\ss}ner-Institut, Bayerische Akademie der Wissenschaften, 85748 Garching, Germany }
\affiliation{Physik-Department, Technische Universit\"{a}t M\"{u}nchen, 85748 Garching, Germany}
\author{F.~Deppe}
\affiliation{Walther-Mei{\ss}ner-Institut, Bayerische Akademie der Wissenschaften, 85748 Garching, Germany }
\affiliation{Physik-Department, Technische Universit\"{a}t M\"{u}nchen, 85748 Garching, Germany}
\affiliation{Nanosystems Initiative Munich (NIM), Schellingstra{\ss}e 4, 80799 M\"{u}nchen, Germany}
\author{K.~G.~Fedorov}
\affiliation{Walther-Mei{\ss}ner-Institut, Bayerische Akademie der Wissenschaften, 85748 Garching, Germany }
\affiliation{Physik-Department, Technische Universit\"{a}t M\"{u}nchen, 85748 Garching, Germany}
\author{P.~Eder}
\affiliation{Walther-Mei{\ss}ner-Institut, Bayerische Akademie der Wissenschaften, 85748 Garching, Germany }
\affiliation{Physik-Department, Technische Universit\"{a}t M\"{u}nchen, 85748 Garching, Germany}
\affiliation{Nanosystems Initiative Munich (NIM), Schellingstra{\ss}e 4, 80799 M\"{u}nchen, Germany}
\author{M.~Fischer}
\affiliation{Walther-Mei{\ss}ner-Institut, Bayerische Akademie der Wissenschaften, 85748 Garching, Germany }
\affiliation{Physik-Department, Technische Universit\"{a}t M\"{u}nchen, 85748 Garching, Germany}
\affiliation{Nanosystems Initiative Munich (NIM), Schellingstra{\ss}e 4, 80799 M\"{u}nchen, Germany}
\author{F.~Wulschner}
\affiliation{Walther-Mei{\ss}ner-Institut, Bayerische Akademie der Wissenschaften, 85748 Garching, Germany }
\affiliation{Physik-Department, Technische Universit\"{a}t M\"{u}nchen, 85748 Garching, Germany}
\author{E.~Xie}
\affiliation{Walther-Mei{\ss}ner-Institut, Bayerische Akademie der Wissenschaften, 85748 Garching, Germany }
\affiliation{Physik-Department, Technische Universit\"{a}t M\"{u}nchen, 85748 Garching, Germany}
\affiliation{Nanosystems Initiative Munich (NIM), Schellingstra{\ss}e 4, 80799 M\"{u}nchen, Germany}
\author{A.~Marx}
\affiliation{Walther-Mei{\ss}ner-Institut, Bayerische Akademie der Wissenschaften, 85748 Garching, Germany }
\author{R.~Gross}
\email[]{rudolf.gross@wmi.badw.de}
\affiliation{Walther-Mei{\ss}ner-Institut, Bayerische Akademie der Wissenschaften, 85748 Garching, Germany }
\affiliation{Physik-Department, Technische Universit\"{a}t M\"{u}nchen, 85748 Garching, Germany}
\affiliation{Nanosystems Initiative Munich (NIM), Schellingstra{\ss}e 4, 80799 M\"{u}nchen, Germany}

\date{\today}

\begin{abstract}
In experiments with superconducting quantum circuits, characterizing the photon statistics of propagating microwave fields is a fundamental task. We quantify the $n^{2}\,{+}\,n$ photon number variance of thermal microwave photons emitted from a black-body radiator for mean photon numbers $0.05\,{\lesssim}\,n\,{\lesssim}\,1.5$.  We probe the fields using either correlation measurements or a transmon qubit coupled to a microwave resonator. Our experiments provide a precise quantitative characterization of weak microwave states and information on the noise emitted by a Josephson parametric amplifier.
\end{abstract}

% \pacs{42.50.Pq,02.50.-r,85.25.Hv}
\keywords{}
\maketitle

As propagating electromagnetic fields in general~\cite{Bouwmeester_1997,Furusawa_1998,Kok_2007}, propagating microwaves with photon numbers on the order of unity are essential for quantum computation~\cite{Braunstein_2005,Andersen_2015}, communication~\cite{Candia_2015}, and illumination~\cite{Lloyd_2008,Tan_2008,Lopaeva_2013,Barzanjeh_2015} protocols. Because of their omnipresence in experimental setups, the characterization of thermal states is especially relevant for many applications~\cite{Yao_2011,Yao_2013,Bohr_2015,Xiang_2016}. Specifically in the microwave regime, sophisticated experimental techniques for their generation at cryogenic temperatures, their manipulation, and detection have been developed in recent years. In this context, an important aspect is the generation of propagating thermal microwaves using thermal emitters~\cite{Gabelli_2004,Mariantoni_2010,Menzel_2010}. These emitters can be spatially separated from the setup components used for manipulation and detection~\cite{Menzel_2012,Fedorov_2016}, which allows one to individually control the emitter and the setup temperature. Due to the low energy of microwave photons, the detection of these fields typically requires the use of near-quantum-limited amplifiers~\cite{Mallet_2011,Zhong_2013,Macklin_2015,Virally_2016}, cross-correlation detectors~\cite{Menzel_2010,Menzel_2012,Eichler_2011}, or superconducting qubits~\cite{Clerk_2007,Sears_2012,Murch_2013,Goetz_2016b}.

\begin{figure}[b]
\includegraphics{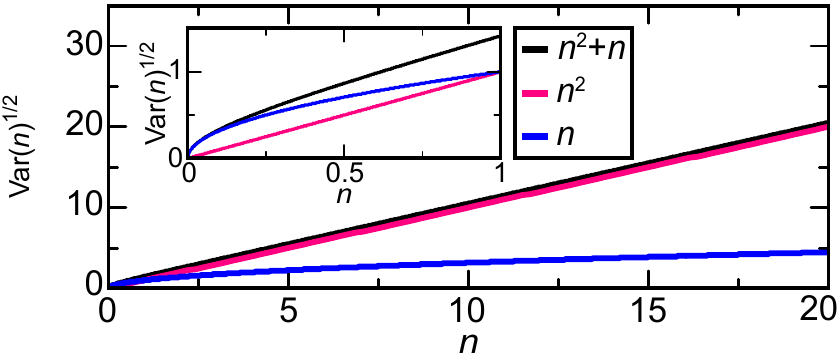}
\caption{\label{fig:scheme} Photon number correlations. [Var$(n)]^{1/2}$ plotted versus photon number for thermal fields (black), their classical limit (red), and coherent states (blue). The inset shows the regime that we capture in our experiments.}
\end{figure}

The unique nature of propagating fields is reflected in their photon statistics, which is described by a probability distribution either in terms of the number states or in terms of its moments. The former were studied by coupling the field to an atom or qubit and measuring the coherent dynamics~\cite{Meekhof_1996,Brune_1996a,Hofheinz_2008} or by spectroscopic analysis~\cite{Schuster_2007a}. The moment-based approach requires knowledge on the average photon number $n$ and its variance Var$(n)\,{=}\,\langle n^{2}\rangle\,{-}\,\langle n\rangle^{2}$ to distinguish many states of interest. To this end, the second-order correlation function $g^{(2)}(\tau)$ has been measured to analyze the photon statistics of thermal~\cite{Morgan_1966,Arecchi_1966,Tan_2014} or quantum~\cite{Short_1983,Rempe_1990,Treussart_2002} emitters ever since the ground-breaking experiments of Hanbury~Brown and Twiss~\cite{Hanbury_19561,Hanbury_1956}. While these experiments use the time delay $\tau$ as control parameter, at microwave frequencies the photon number $n$ can be controlled conveniently~\cite{Gabelli_2004,Schuster_2005,Schuster_2007a,Houck_2007,Fink_2010,Forgues_2014}. In the specific case of a thermal field at frequency $\omega$, the Bose-Einstein distribution yields $n(T)\,{=}\,[\exp(\hbar\omega/k_{\mathrm{B}}T)\,{-}\,1]^{-1}$ and Var$(n)\,{=}\,n^{2}\,{+}\,n$, which can be controlled by the temperature $T$ of the emitter. In practice, one wants to distinguish this relation from both the classical limit Var$(n)\,{=}\,n^{2}$ and the Poissonian behavior Var$(n)\,{=}\,n$ characteristic for coherent states~\cite{Schuster_2005} or shot noise~\cite{Beenakker_1999,Blanter_2000}. Hence, as shown in Fig.\,\ref{fig:scheme}, the most relevant regime for experiments is $n\,{\lesssim}\,1$, which translates into temperatures between \SI{100}{\milli\kelvin} and \SI{1}{\kelvin} at approximately \SI{6}{\giga\hertz} for the thermal emitter~\cite{Goetz_2016b}.

In this Letter, we experimentally confirm the theoretically expected photon number variance Var$(n)$ of thermal microwave fields for $n\,{\lesssim}\,1.5$ using two fundamentally distinct experimental setups. To this end, we first use a superconducting transmon qubit~\cite{Koch_2007} interacting with the propagating fields via a dispersively coupled microwave resonator. Differently to approaches relying on the coherent dynamics~\cite{Meekhof_1996,Brune_1996a,Hofheinz_2008}, where decoherence is detrimental, the additional qubit dephasing rate induced by the field directly reflects the photon number variance in our experiments. We furthermore get access to finite-time correlations for fields with Poissonian photon statistics because the resonator mediates different decay constants for the photon-photon correlator of incoherent and coherent noise. In particular, we find the expected factor of two between the dephasing rates caused by coherent states and shot noise. With the second setup, we extract the super-Poissonian photon statistics of propagating thermal microwaves from direct correlation measurements and from measurements using a near-quantum-limited Josephson parametric amplifier (JPA)~\cite{Yamamoto_2008,Zhong_2013} as preamplifier. The results show that the noise added by the JPA inevitably alters the photon statistics of the amplified field. Our results provide a quantitative picture of propagating thermal microwaves, which is especially relevant for the characterization of more advanced quantum states in the presence of unavoidable thermal background fields. With respect to superconducting qubits, we gain systematic insight into a dephasing mechanism which may become relevant for state-of-the-art devices with long coherence times~\cite{Rigetti_2012,Yan_2015}.

In our experiments, we generate the thermal fields using a temperature-controllable, \SI{50}{\ohm}-matched attenuator acting as a black-body emitter. This emitter is thermally only weakly coupled to the \SI{35}{\milli\kelvin} base temperature stage of a dilution refrigerator. Heating the attenuator up to \SI{1.5}{\kelvin} results in the emission of thermal microwave radiation with a photon number stability $[\text{Var}(n)]^{1/2}/n\,{\lesssim}\,0.01$. In addition, we investigate coherent states emitted from a microwave source and white electronic shot noise with a \SI{200}{\mega\hertz} bandwidth generated by an arbitrary function generator (AFG). The AFG output is upconverted to a center frequency of \SI{6.05}{\giga\hertz} (see Ref.~\footnote{See Supplemental Material [url], which includes Refs.~\onlinecite{Kano_1962,Mehta_1963,Lindblad_1976,Caves_1982,Gambetta_2008,Klein_2008,Boissonneault_2009,Clerk_2010,Devoret_2014,Kochetov_2015,Goetz_2016}.} for details). For coherent states and shot noise, the photon number entering the cryostat is proportional to the power set at the microwave source or the AFG, respectively.

\begin{figure}[t]
\includegraphics{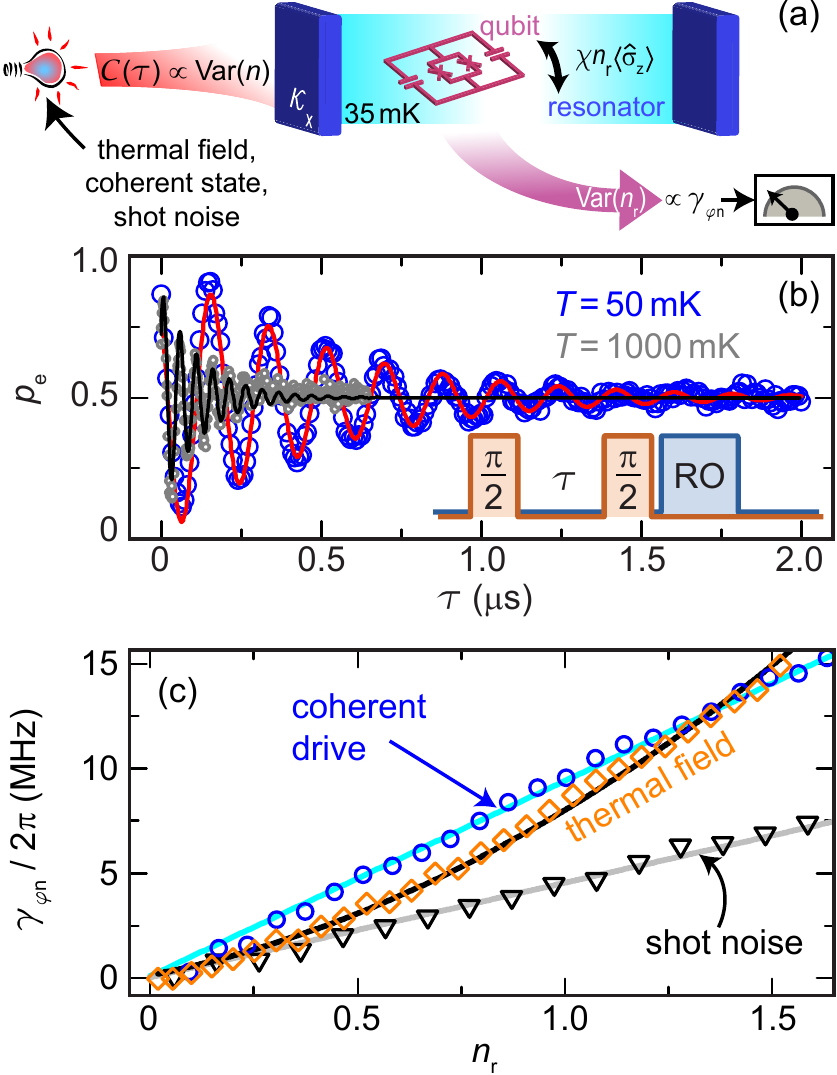}
\caption{\label{fig:T2} (a) Sketch of the qubit setup. We measure the photon number variance Var$(n)$ of microwave fields encoded in the photon correlator~$\mathcal{C}(\tau)$ by detecting the dephasing rate~$\gamma_{\varphi n}$ of a superconducting qubit. (b) Qubit excited state probability~$p_{\mathrm{e}}$ for a Ramsey experiment plotted versus waiting time $\tau$ between two~$\pi/2$ pulses. The solid lines are exponentially decaying sinusoidal fits. Inset: Ramsey pulse sequence followed by a readout (RO) pulse. (c) Qubit dephasing rates~$\gamma_{\varphi n}$ of prototypical input fields plotted versus the average resonator population $n_{\mathrm{r}}$, which is calibrated in an AC Stark shift measurement~\cite{Note1}. Solid lines are fits using Eqs.\,(\ref{eqn:gammath})\,-\,(\ref{eqn:gammash}).}
\end{figure}

To measure the photon number fluctuations of propagating microwaves, we enhance their lifetime by trapping them inside a coplanar waveguide resonator. The latter is dispersively coupled to a superconducting transmon qubit acting as a sensitive detector [see Fig.\,\ref{fig:T2}(a)]. The transmon qubit is frequency-tunable and operated at its maximum transition frequency~$\omega_{\mathrm{q}}/2\pi\,{=}\,\SI{6.92}{\giga\hertz}$. The resonator with resonance frequency~$\omega_{\mathrm{r}}/2\pi\,{=}\,\SI{6.07}{\giga\hertz}$ is characterized by its external coupling and internal loss rate  $\kappa_{\mathrm{x}}/2\pi\,{=}\,\SI{8.5}{\mega\hertz}$ and $\kappa_{\mathrm{i}}/2\pi\,{\simeq}\,\SI{50}{\kilo\hertz}$. The dispersive interaction Hamiltonian reads $\mathcal{H}_{\mathrm{int}}\,{=}\,\hbar\chi[n_{\mathrm{r}}\,{+}\,1/2]\hat{\sigma}_{\mathrm{z}}$, where~\cite{Koch_2007}~$\chi\,{\equiv}\,[g^{2}/\delta][\alpha/(\delta\,{+}\,\alpha)]\,{\simeq}\,{-}2\pi\,{\times}\,\SI{3.11}{\mega\hertz}$. In this expression, $g/2\pi\,{\simeq}\,\SI{67}{\mega\hertz}$ is the qubit-resonator coupling, $\alpha/2\pi\,{\simeq}\,{-}\SI{315}{\mega\hertz}$ is the transmon anharmonicity, and $\delta\,{\equiv}\,\omega_{\mathrm{q}}\,{-}\,\omega_{\mathrm{r}}$ is the detuning. Following input-output formalism~\cite{Gardiner_1994,Ridolfo_2012}, the photon number fluctuations $n(\tau)$ of the incident fields have the same statistics as $n_{\mathrm{r}}(\tau)$ for our sample parameters. Because these fluctuations couple to the qubit via the Pauli operator $\hat{\sigma}_{\mathrm{z}}$, they introduce qubit dephasing characterized by the photon-photon correlator $\mathcal{C}(\tau)\,{\equiv}\,\langle n_{\mathrm{r}}(0)n_{\mathrm{r}}(\tau)\rangle\,{-}\,\langle n_{\mathrm{r}}(0)\rangle^{2}$~\cite{Blais_2004}. For all microwave states discussed here, $\mathcal{C}(\tau)\,{=}\,\mathrm{Var}(n_{\mathrm{r}})\exp({-}\tilde{\kappa}\tau)$ factorizes into the photon number variance and a temporal decay with rate $\tilde{\kappa}$ due to the resonator~\cite{Note1}. For incoherent signals with white spectrum, $\tilde{\kappa}\,{=}\,\kappa_{\mathrm{x}}$ corresponds to the energy decay rate of the resonator. Nevertheless, the thermal correlator $\mathcal{C}^{\mathrm{th}}(\tau)\,{=}(n_{\mathrm{r}}^{2}\,{+}\,n_{\mathrm{r}})\exp({-}\kappa_{\mathrm{x}}\tau)$ can be distinguished from the shot noise correlator $\mathcal{C}^{\mathrm{sh}}(\tau)\,{=}\,n_{\mathrm{r}}\exp({-}\kappa_{\mathrm{x}}\tau)$ via their photon number variance. Remarkably, despite sharing a Poissonian photon statistics, $\mathcal{C}^{\mathrm{sh}}(\tau)$ differs from the coherent state correlator $\mathcal{C}^{\mathrm{coh}}(\tau)\,{=}\,n_{\mathrm{r}}\exp({-}\kappa_{\mathrm{x}}\tau/2)$, because the latter decays at the amplitude decay rate $\tilde{\kappa}\,{=}\,\kappa_{\mathrm{x}}/2$. All three correlators generate a shift $\delta\varphi(\tau)$ of the qubit phase, whose second moment~\cite{Gambetta_2006} $\langle\delta\varphi^{2}\rangle\,{=}\,4\chi^{2}
\int_{0}^{\tau}\mathrm{d}\tau^{\prime}\mathcal{C}(\tau^{\prime})$ enters into the Ramsey decay envelope $\exp[{-}\gamma_{1}(n_{\mathrm{r}})\tau/2\,{-}\,\gamma_{\varphi0}\tau\,{-}\langle\delta\varphi^{2}\rangle/2]$. Here, $\gamma_{1}(n_{\mathrm{r}})$ is the total qubit relaxation rate and $\gamma_{\varphi0}$ is the qubit dephasing rate due to all other noise sources except for those described by $\mathcal{C}(\tau)$. Assuming $\langle\delta\varphi^{2}\rangle/2\,{=}\,\gamma_{\varphi n}(n_{\mathrm{r}})\tau$ and $|\chi|\,{\ll}\,\kappa_{\mathrm{x}}$, the photon-field-induced dephasing rates approximate to~\cite{Schuster_2005,Gambetta_2006,Clerk_2007,Rigetti_2012,Bertet_2005,Bertet_2005a}
\begin{eqnarray}
\gamma_{\varphi n}^{\mathrm{th}}(n_{\mathrm{r}}) =& \kappa_{\mathrm{x}}\theta_{0}^{2}(n_{\mathrm{r}}^{2}+n_{\mathrm{r}}) &\equiv s_{0}^{\mathrm{th}}(n_{\mathrm{r}}^{2}+n_{\mathrm{r}})\,,\label{eqn:gammath}\\
\gamma_{\varphi n}^{\mathrm{coh}}(n_{\mathrm{r}}) =& 2\kappa_{\mathrm{x}}\theta_{0}^{2}n_{\mathrm{r}} &\equiv s_{0}^{\mathrm{coh}}n_{\mathrm{r}}\,,  \label{eqn:gammacoh}\\
\gamma_{\varphi n}^{\mathrm{sh}}(n_{\mathrm{r}}) =& \kappa_{\mathrm{x}}\theta_{0}^{2}n_{\mathrm{r}} &\equiv s_{0}^{\mathrm{sh}}n_{\mathrm{r}}\,.  \label{eqn:gammash}
\end{eqnarray}
Here, $\theta_{0}\,{\equiv}\,\tan^{-1}(2\chi/\kappa_{\mathrm{x}})$ is the accumulated phase of the resonator photons due to the interaction with the qubit. The factor two between $\gamma_{\varphi n}^{\mathrm{coh}}$ and $\gamma_{\varphi n}^{\mathrm{sh}}$ reflects the fact that the impact of the fluctuations onto the qubit is larger if the correlator decays slower. As a consequence of Eqs.\,(\ref{eqn:gammath})\,-\,(\ref{eqn:gammash}), measurements of the Ramsey decay rate $\gamma_{2}(n_{\mathrm{r}})\,{=}\,\gamma_{1}(n_{\mathrm{r}})/2\,{+}\,\gamma_{\varphi0}\,{+}\,\gamma_{\varphi n}(n_{\mathrm{r}})$ allow us to extract Var$(n_{\mathrm{r}})$ after correcting $\gamma_{2}(n_{\mathrm{r}})$ for $\gamma_{1}(n_{\mathrm{r}})$ obtained from an independent measurement~\cite{Note1,Goetz_2016b}. We emphasize that during our sweeps of the attenuator temperature, the sample box is stabilized at $\SI{35}{\milli\kelvin}$. Therefore, $\gamma_{\varphi0}$ can be taken as a constant and we can extract $\gamma_{\varphi n}$ from the decay envelope of a Ramsey time trace.

In the absence of external microwave fields, the transmon qubit is relaxation-limited with the rates $\gamma_{1}(n_{\mathrm{r}}\,{\simeq}\,0)/2\pi\,{\simeq}\,\SI{4}{\mega\hertz}$ and $\gamma_{2}(n_{\mathrm{r}}\,{\simeq}\,0)/2\pi\,{\simeq}\,\SI{2}{\mega\hertz}$. In Fig.\,\ref{fig:T2}(b), we show the Ramsey time traces for the attenuator temperatures $T\,{=}\,\SI{50}{\milli\kelvin}$ and $T\,{=}\,\SI{1}{\kelvin}$. As expected, the latter shows a significantly increased Ramsey decay rate. A systematic temperature sweep reveals $\gamma_{\varphi n}^{\mathrm{th}}(n_{\mathrm{r}})\,{\propto}\,n_{\mathrm{r}}^{2}\,{+}\,n_{\mathrm{r}}$ as displayed in Fig.\,\ref{fig:T2}(c). For small photon numbers $n_{\mathrm{r}}\,{\lesssim}\,0.5$, the dephasing rate approaches a linear trend with slope $s_{0}^{\mathrm{th}}\,{\equiv}\,\partial\gamma_{\varphi n}^{\mathrm{th}}/\partial n_{\mathrm{r}}|_{n_{\mathrm{r}}\,{=}\,0}$. This finite slope clearly allows us to rule out the validity of the classical limit Var$(n_{\mathrm{r}})\,{=}\,n_{\mathrm{r}}^{2}$ in this regime. From a fit of Eq.\,(\ref{eqn:gammath}) to the data, we find~$s_{0}^{\mathrm{th}}/2\pi\,{=}\,\SI{3.9}{\mega\hertz}$, which is marginally enhanced compared to the expected value $\kappa_{\mathrm{x}}\theta_{0}^{2}/2\pi\,{=}\,\SI{3.4}{\mega\hertz}$. Because the enhancement of $s_{0}^{\mathrm{th}}$ cannot be linked to the finite cavity pull $|\chi/\kappa_{\mathrm{x}}|\,{\simeq}\,0.3$, we attribute it to thermal photons $n_{\mathrm{n}}^{\mathrm{th}}$ emitted from attenuators at higher temperature stages~\cite{Note1}. Applying a beam splitter model to calculate Var$(n_{\mathrm{r}}\,{+}\,n_{\mathrm{n}}^{\mathrm{th}})$ yields the reasonable contribution of $n_{\mathrm{n}}^{\mathrm{th}}\,{=}\,0.15$ corresponding to an effective mode temperature of approximately \SI{140}{\milli\kelvin}.

As a cross-check for our setup, we confirm the well-explored~\cite{Schuster_2005,Gambetta_2006,Sears_2012,Virally_2016} linear variance of fields with Poissonian photon statistics. To this end, we first expose the resonator to shot noise emitted at room temperature by the AFG. As shown in Fig.\,\ref{fig:T2}(c), we indeed find a constant slope $s^{\mathrm{sh}}\,{\equiv}\,\partial\gamma_{\varphi n}^{\mathrm{sh}}/\partial n_{\mathrm{r}}\,{\simeq}\,2\pi\,{\times}\,\SI{4.6}{\mega\hertz}$, which is in reasonable agreement with $s_{0}^{\mathrm{th}}$. In terms of additional thermal population and effective mode temperature, we obtain $n_{\mathrm{n}}^{\mathrm{sh}}\,{\simeq}\,0.19\,{\approx}\,n_{\mathrm{n}}^{\mathrm{th}}$ and \SI{150}{\milli\kelvin}, respectively. In the next step, we investigate measurement-induced dephasing caused by coherent states. We again find a linear slope $s^{\mathrm{coh}}\,{\equiv}\,\partial\gamma_{\varphi n}^{\mathrm{coh}}/\partial n_{\mathrm{r}}\,{\simeq}\,2\pi\,{\times}\,\SI{9.3}{\mega\hertz}$. Although both coherent states and shot noise exhibit Poissonian statistics, we can reliably distinguish between the two of them using the fact that $s^{\mathrm{coh}}\,{\simeq}\,2s^{\mathrm{sh}}$. This discrimination shows that the qubit dephasing rate directly reflects the temporal dependence of photon-photon correlators. The excellent quantitative agreement is also reflected in $n_{\mathrm{n}}^{\mathrm{coh}}\,{=}\,n_{\mathrm{n}}^{\mathrm{sh}}$, i.e., identical Fano factors~\cite{Beenakker_1999} $\mathcal{F}\,{\equiv}\,\mathrm{Var}(n_{\mathrm{r}})/n_{\mathrm{r}}\,{\simeq}\,1.1$.

\begin{figure}[t]
\includegraphics{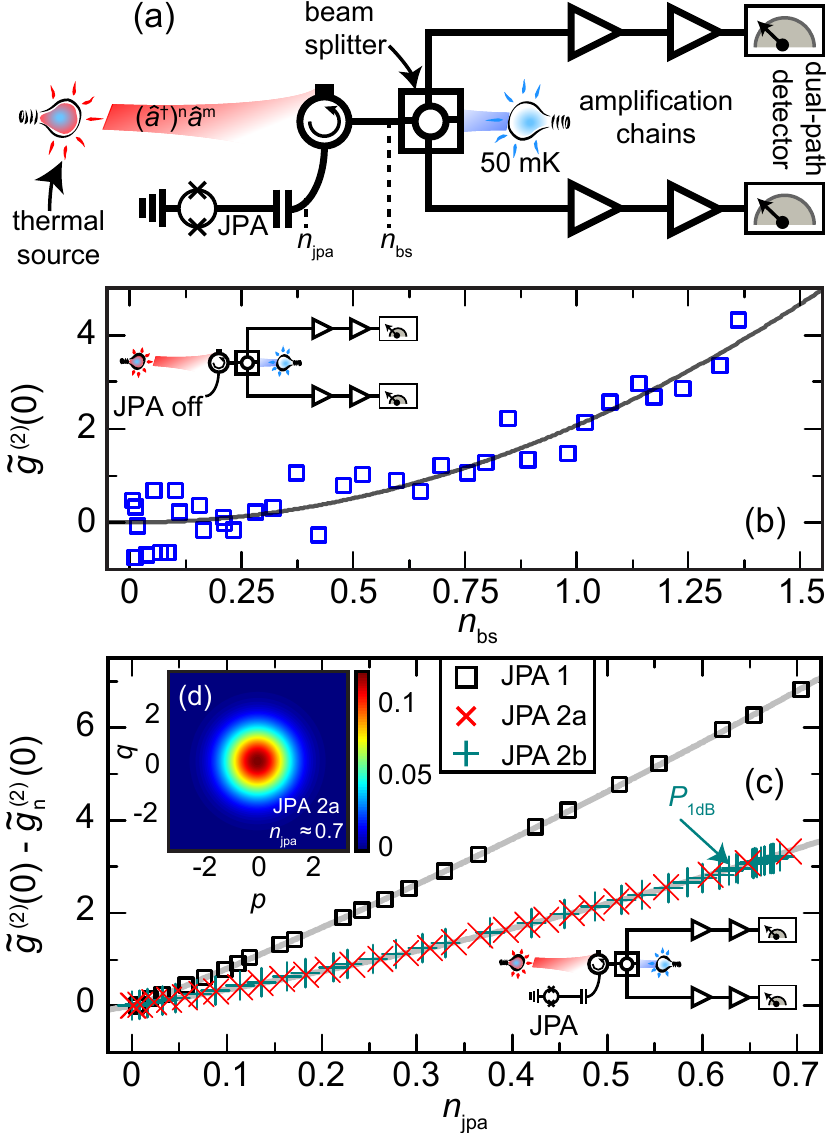}
\caption{\label{fig:crosscorr} (a) Sketch of the dual-path setup, which we use to probe field correlations between two amplification chains behind a cryogenic microwave beam splitter. We can switch the JPA on and off. (b) Unnormalized second-order correlation function $\tilde{g}^{(2)}(0)$ plotted versus photon number $n_{\mathrm{bs}}$ at the beam splitter input without using the JPA. The solid line is a fit to the data using the function $\tilde{g}^{(2)}(0)\,{=}\,\rho\,n_{\mathrm{bs}}^{2}$. (c) Unnormalized second-order correlation function $\tilde{g}^{(2)}(0)$ corrected for the constant offset $\tilde{g}^{(2)}_{\mathrm{n}}(0)$ and plotted versus the photon number $n_{\mathrm{jpa}}$ at the JPA input. For the measurements of JPA\,2a and JPA\,2b we use slightly different operating points described in detail in Ref.\,\onlinecite{Note1}. (d) Wigner function reconstruction referred to the input of the JPA for a thermal state.}
\end{figure}

In order to complement our studies of thermal microwaves, we directly probe field correlations with the dual-path state reconstruction method~\cite{Mariantoni_2010,Menzel_2010,Menzel_2012,Candia_2014}. This approach is motivated by the prediciton that a beam splitter transfers the photon statistics of two uncorrelated inputs into correlations between its two outputs~\cite{Campos_1989}. We use the setup depicted in Fig.\,\ref{fig:crosscorr}(a), where a cryogenic beam splitter equally divides the signal along two paths, which are subsequently amplified independently. From their averaged auto- and cross-correlations, we retrieve all signal moments $\langle(\hat{a}^{\dagger})^{n}\hat{a}^{m}\rangle$ up to fourth order ($0\,{\leq}\,n\,{+}\,m\,{\leq}\,4$ with $n,m\,{\in}\,\mathbb{N}_{0}$) in terms of the annihilation and creation operators, $\hat{a}$ and $\hat{a}^{\dagger}$. To calibrate the average photon number $n_{\mathrm{bs}}\,{=}\,\langle\hat{a}^{\dagger}\hat{a}\rangle\,{\propto}\,n(T)$ at the input of the beam splitter, we perform a Planck spectroscopy experiment~\cite{Mariantoni_2010} (see Ref.~\onlinecite{Note1} for details). Notwithstanding the very different experimental requirements in the microwave regime, direct correlation measurements on propagating light fields are inspired from quantum optics. For this reason, we characterize the photon number variance of the thermal microwave fields via the unnormalized correlation function
\begin{equation}
\tilde{g}^{(2)}(0) \equiv n_{\mathrm{bs}}^{2}g^{(2)}(0) = \mathrm{Var}(n_{\mathrm{bs}}^{})\,{-}\,n_{\mathrm{bs}}^{}\,{+}\,n_{\mathrm{bs}}^{2}\,.
\label{eqn:g2}
\end{equation}
As shown in Fig.\,\ref{fig:crosscorr}(b), the correlation function $\tilde{g}^{(2)}(0)$ of the thermal source follows the expected quadratic behavior. A numerical fit of the polynomial function $\tilde{g}^{(2)}(0)\,{=}\,\rho\,n_{\mathrm{bs}}^{2}$ using $\rho$ as a free parameter yields $\rho\,{=}\,2.07$. This result coincides nicely with $\tilde{g}^{(2)}(0)\,{=}\,2n_{\mathrm{bs}}^{2}$ predicted for thermal states by Eq.\,(\ref{eqn:g2}). In the same way as with the qubit setup, we are therefore able to reliably map out the $n^{2}\,{+}\,n$ dependence and not only the classical $n^{2}$ limit experimentally found in earlier work~\cite{Gabelli_2004}.

To lower the statistical scatter of the data points in Fig.\,\ref{fig:crosscorr}(b), we repeat the correlation measurement using a JPA operated in the phase-insensitive mode. In this mode, the JPA works as a near-quantum-limited, phase-preserving amplifier~\cite{Zhong_2013} with power gain $G\,{\gg}\,1$. At the input of the beam splitter, one then obtains $n_{\mathrm{bs}}\,{\approx}\,G(n_{\mathrm{jpa}}\,{+}\,n_{\mathrm{n}}\,{+}\,1)$. Here, $n_{\mathrm{jpa}}\,{\propto}\,n(T)$ are the signal photons and $n_{\mathrm{n}}$ are the noise photons added by the JPA, which we again obtain from a Planck spectroscopy experiment~\cite{Note1}. We compare measurements using two different JPAs (JPA\,1 and JPA\,2) based on frequency-tunable quarter wavelength resonators with operating frequencies~$\omega_{\mathrm{jpa}}/2\pi\,{\simeq}\,\SI{5.35}{\giga\hertz}$ and typical gains $G\,{\simeq}\,\SI{15}{\decibel}$. To characterize the noise referred to the input of the JPA, we analyze the modified correlation function
\begin{equation}
\tilde{g}^{(2)}(0) = 2(n_{\mathrm{jpa}}\,{+}\,n_{\mathrm{n}}\,{+}\,1)^{2}\,,
\label{eqn:g2jpa}
\end{equation}
which can be derived from an input-output model for the JPA. In our model, we assume that the JPA noise is thermal, i.e., $\mathrm{Var}(n_{\mathrm{n}})\,{=}\,n_{\mathrm{n}}^{2}\,{+}\,n_{\mathrm{n}}$. Then, there is a $n_{\mathrm{jpa}}\,{-}\,$independent offset $\tilde{g}^{(2)}_{\mathrm{n}}(0)\,{=}\,2n_{\mathrm{n}}^{2}\,{+}\,4n_{\mathrm{n}}\,{+}\,2$ in Eq.\,(\ref{eqn:g2jpa}) due to the JPA gain and noise.

In Fig.\,\ref{fig:crosscorr}(c) we plot the experimentally obtained correlations $\tilde{g}^{(2)}(0)\,{-}\,\tilde{g}^{(2)}_{\mathrm{n}}(0)$ versus the photon number $n_{\mathrm{jpa}}$ at the JPA input. From fits to the formula $\rho n_{\mathrm{jpa}}^2\,{+}\,\xi n_{\mathrm{jpa}}$, we find $\rho\,{\simeq}\,2.2$ in all three data sets in agreement with the expected value of $\rho\,{=}\,2$. Therefore, also the JPA assisted measurements confirm a super-Poissonian statistics of the thermal fields. From the fits, we also find that the values of $\xi$ are reduced by a factor of approximately $2$ compared to the expected value $4\,{+}\,4n_{\mathrm{n}}$. This observation is confirmed by the values extracted for $\tilde{g}^{(2)}_{\mathrm{n}}(0)$, which deviate to a similar extent. Assuming that the photon statistics of the signal photons $n_{\mathrm{jpa}}$ is not changed by the JPA, the reduced experimental values suggest that the amplified noise contains a significant contribution Var$(n_{\mathrm{n}})\,{=}\,n_{\mathrm{n}}^{2}$. This classical contribution is power independent and unaffected when the JPAs exceed their \SI{1}{\decibel} compression point $\mathcal{P}_{1\mathrm{dB}}\,{\simeq}\,\SI{-130}{\decibel}$m [see Fig.\,\ref{fig:crosscorr}(c)]. We stress that the amplified fields are still Gaussian and show no squeezing effects between the two quadratures $\hat{p}\,{=}\,\imath(\hat{a}^{\dagger}\,{-}\,\hat{a})/2$ and $\hat{q}\,{=}\,(\hat{a}^{\dagger}\,{+}\,\hat{a})/2$ [see Fig.\,\ref{fig:crosscorr}(d)]. As shown in Ref.~\onlinecite{Note1}, we find  Var$(\hat{p})\,{=}\,\mathrm{Var}(\hat{q})$ for the complete temperature range.

Finally, we compare the performance of the qubit and the dual-path setup. Although we operate on and below the single photon level, the qubit and the dual-path setup (without JPA) systematically reproduce the $n^{2}\,{+}\,n$ law with a high accuracy. Currently, the statistical spread for the qubit setup is one order of magnitude lower than the one for the dual-path setup. The accuracy of the qubit setup is limited by the Fano factor $\mathcal{F}\,{\simeq}\,1.1$ of the setup and by the low-frequency variations of the qubit relaxation rate described in Ref.~\onlinecite{Goetz_2016b}. Their standard deviation of 5\,\% well explains the spread of the experimental data points in Fig.\,\ref{fig:T2}(c). Assuming that these variations decrease proportionally to the qubit decoherence rate, we estimate that for the best performing superconducting qubits~\cite{Rigetti_2012}, the accuracy can be improved by at least two orders of magnitude. The dual-path setup (without the JPA) is limited by the data processing rate of our digitizer card and by the noise temperature $T_{\mathrm{n}}\,{\simeq}\,\SI{3}{\kelvin}$ of the cryogenic amplifiers. When the JPA is on, the noise temperature of these amplifiers is insignificant. While our measurements including a JPA decrease the statistical spread by two orders of magnitude, they also introduce a systematic error due to uncertainties in the JPA noise statistics. Concerning adaptability, the dual-path setup in principle gives access to all signal moments, whereas the qubit is limited to amplitude and power correlations. 

In conclusion, we have quantitatively characterized the photon number variance of propagating thermal microwaves using two fundamentally different approaches: indirect measurements with a superconducting qubit-resonator system and direct ones, with a dual-path detector. With both setups, we are able to quantitatively recover the $n^{2}\,{+}\,n$ photon number variance of thermal fields in the single photon regime with a high resolution in comparison with existing experimental achievements~\cite{Gabelli_2004}. In particular, we analyze the resolution limits and find that they may improve by several orders of magnitude in both setups. For our current dual-path setup, we make the remarkable observation that noise added by the JPAs has a significant contribution with Var$(n)\,{=}\,n^{2}$. Our results demonstrate that the three types of propagating microwave states we investigate are reliably distinguishable below the single photon level in an experiment by their photon statistics. Therefore, both setups are promising candidates to explore decoherence mechanisms possibly limiting high-performance superconducting qubits~\cite{Rigetti_2012,Yan_2015} and the properties of more advanced quantum microwave states.

The JPAs used in this work are kindly provided by K.~Inomata (RIKEN Center for Emergent Matter Science), T.~Yamamoto (NEC IoT Device Research Laboratories), and Y.~Nakamura (RIKEN, RCAST at the University of Tokyo). We thank E.~Solano, R.~Di~Candia, M.~Sanz (Department of Physical Chemistry, University of the Basque Country) for fruitful discussions on the correlation measurement setup. We acknowledge financial support from the German Research Foundation through SFB 631 and FE 1564/1-1, EU projects CCQED, PROMISCE, the doctorate program ExQM of the Elite Network of Bavaria, and the International Max Planck Research School "Quantum Science and Technology``.

\bibliography{D:/Dropbox/Goetz_Bibliography}

%%%%%%%%%% Merge with supplemental materials %%%%%%%%%%
\newpage
\widetext
\begin{center}
\textbf{\large Supplemental Materials: Photon Statistics of Propagating Thermal Microwaves}
\end{center}
%%%%%%%%%% Merge with supplemental materials %%%%%%%%%%
%%%%%%%%%% Prefix a "S" to all equations, figures, tables and reset the counter %%%%%%%%%%
\setcounter{equation}{0}
\setcounter{figure}{0}
\setcounter{table}{0}
\makeatletter
\renewcommand{\theequation}{S\arabic{equation}}
\renewcommand{\thefigure}{S\arabic{figure}}
\renewcommand{\bibnumfmt}[1]{[S#1]}

\section{Qubit-resonator setup and generation of broadband noise}

\begin{figure*}
\includegraphics{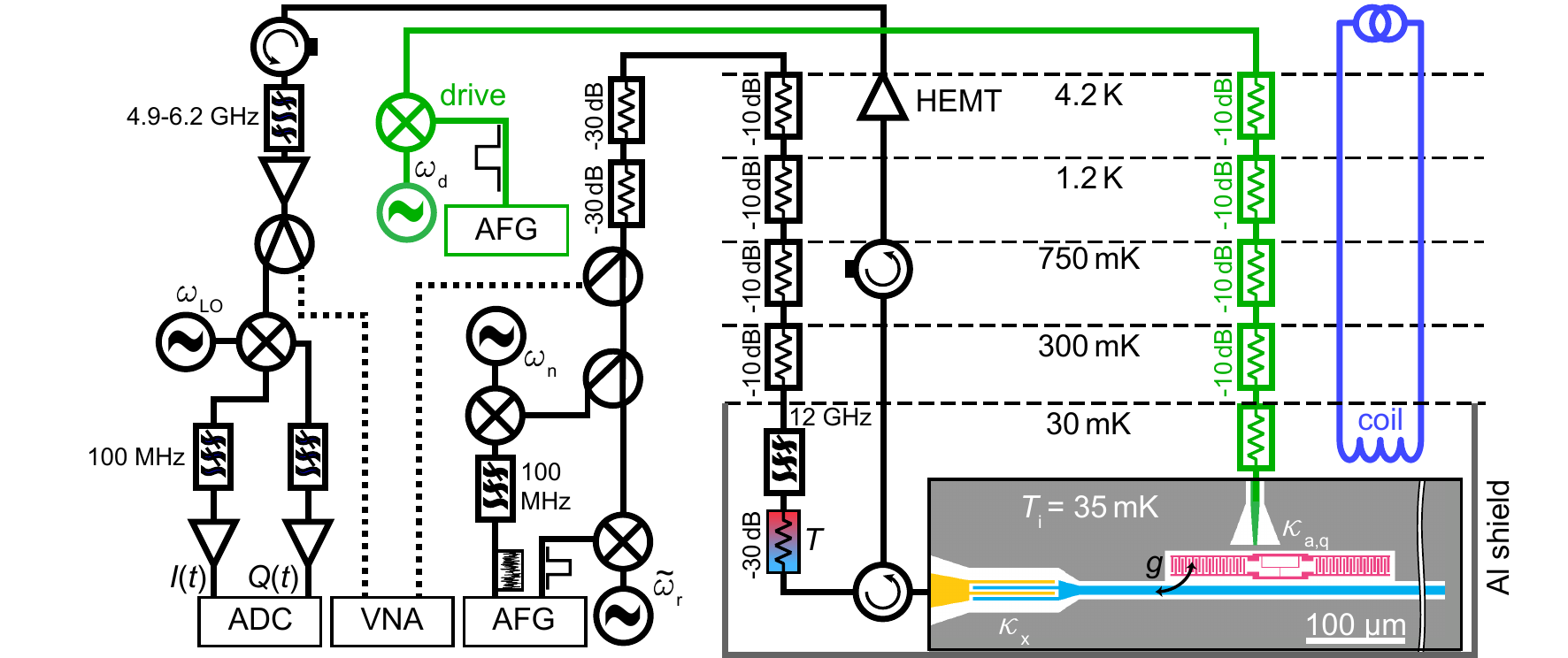}
\caption{\label{fig:S01}Experimental setup: The frequency-tunable transmon qubit is capacitively coupled with coupling strength~$g$ to a readout resonator, which itself is coupled with~$\kappa_{\mathrm{x}}$ to a readout line. Furthermore, the qubit is coupled with~$\kappa_{\mathrm{a,q}}$ to an on-chip antenna. We inject thermal states into the resonator by controlling the temperature~$T$ of a heatable attenuator while keeping the sample temperature constant at~$T_{\mathrm{i}}\,{=}\,\SI{35}{\milli\kelvin}$. Shot noise is generated by upconversion of the noise emitted by an AFG.}
\end{figure*}

\paragraph{Qubit-resonator sample} For the qubit-resonator sample, we use a frequency-tunable superconducting transmon qubit dispersively coupled to a quarter-wavelength coplanar waveguide resonator [see Fig.\,\ref{fig:S01}]. We mount the sample chip on a copper plated printed circuit board inside a gold plated sample box, which is placed on the base temperature stage of a dilution refrigerator. We stabilize the sample box to \SI{35.0+-0.1}{\milli\kelvin}. The qubit is characterized by the $T_{1}$-limited coherence time of approximately \SI{500}{\nano\second} and the energy ratio $E_{\mathrm{J0}}/E_{\mathrm{C}}\,{\simeq}\,64$. Here,~$E_{\mathrm{C}}\,{\simeq}\,h\,{\times}\,\SI{315}{\mega\hertz}$ is the charging energy of the transmon qubit and~$E_{\mathrm{J0}}\,{\simeq}\,h\,{\times}\,\SI{20}{\giga\hertz}$ is the total Josephson energy of the two SQUID junctions, which are used to tune the qubit transition frequency. The magnetic flux~$\Phi$ in the SQUID loop is induced via a superconducting coil outside the sample holder. The qubit is made from a \SI{110}{\nano\meter}-thick Al/AlO$_{\mathrm{x}}$/Al trilayer structure, shadow evaporated onto an undoped Si substrate. The Si is covered with \SI{50}{\nano\meter} thermal oxide on each side resulting in a resistivity larger than \SI{1}{\kilo\ohm\centi\meter} at room temperature. The qubit is coupled with coupling strength~$g/2\pi\,{\simeq}\,\SI{67}{\mega\hertz}$ to the resonator, which we fabricate using optical lithography from a \SI{100}{\nano\meter}-thick Nb film~\cite{Goetz_2016}. The \SI{50}{\ohm}-matched resonator has a resonance frequency~$\omega_{\mathrm{r}}/2\pi\,{\simeq}\,\SI{6.07}{\giga\hertz}$ and is characterized by the internal loss rate~$\kappa_{\mathrm{i}}/2\pi\,{\simeq}\,\SI{50}{\kilo\hertz}$ and the external loss rate~$\kappa_{\mathrm{x}}/2\pi\,{\simeq}\,\SI{8.5}{\mega\hertz}$. In addition to the resonator, the qubit is coupled with coupling rate~$\kappa_{\mathrm{a,q}}/2\pi\,{\simeq}\,\SI{820}{\kilo\hertz}$ to a broadband on-chip antenna, which itself is coupled with~$\kappa_{\mathrm{a,r}}/2\pi\,{\simeq}\,\SI{30}{\kilo\hertz}$ to the resonator.\medskip

\begin{figure*}
\includegraphics{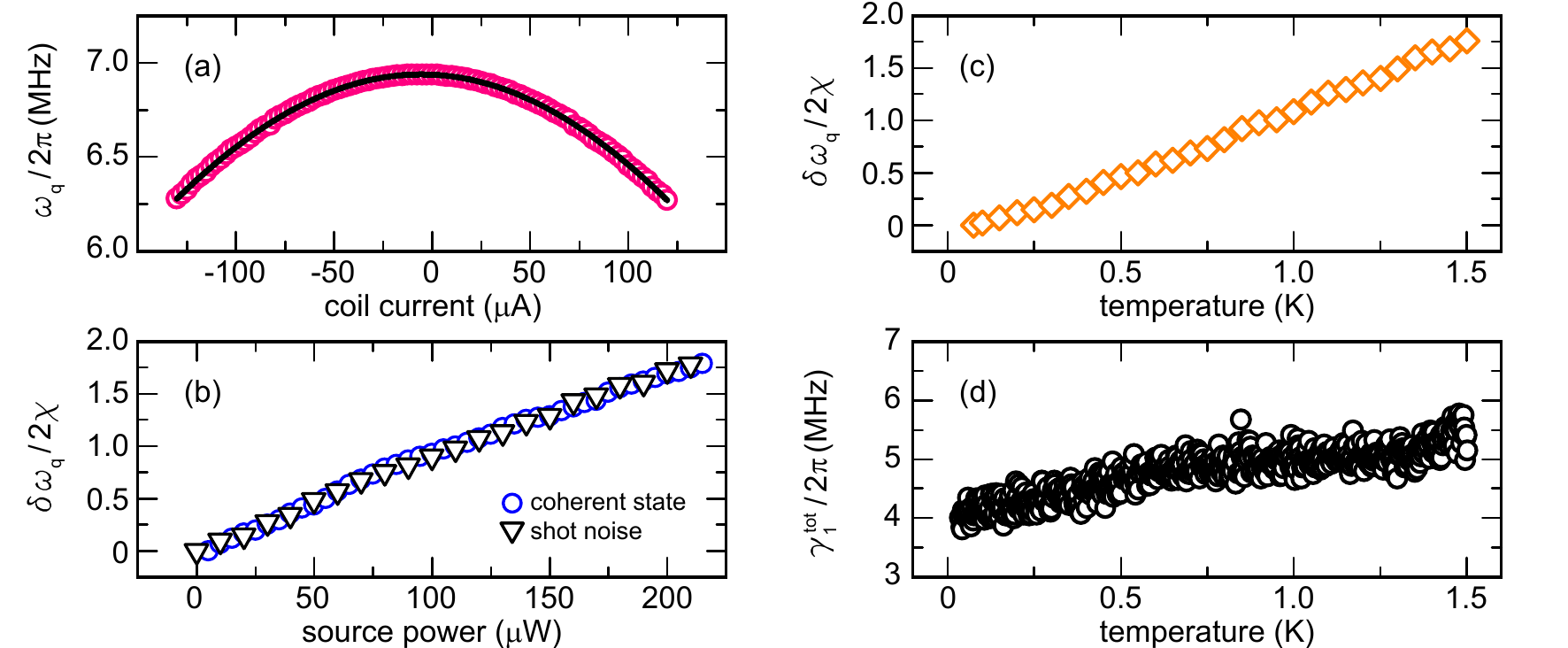}
\caption{\label{fig:S02} (a) Qubit transition frequency $\omega_{\mathrm{q}}$ plotted versus DC current fed into a superconducting coil to generate a magnetic field $\Phi$. The solid line is a numerical fit. (b) Shift of the qubit transition frequencies $\delta\omega_{\mathrm{q}}$ plotted versus the average power of a coherent microwave tone or upconverted shot noise. (c) Shift of the qubit transition frequencies $\delta\omega_{\mathrm{q}}$ plotted versus temperature $T$ of the blackbody emitter. (d) Total qubit relaxation rate $\gamma_{1}^{\mathrm{tot}}$ plotted versus temperature $T$ of a blackbody emitter.}
\end{figure*}

\paragraph{Measurement setup} In our experiments, we perform an averaged weak dispersive readout to detect the qubit state~\cite{Schuster_2005}. We implement a time-resolved, phase-sensitive measurement of the in-phase and quadrature components~$I(t)$ and~$Q(t)$ of the readout signal by heterodyne down conversion of the readout signal at $\tilde{\omega}_{\mathrm{r}}\,{=}\,\omega_{\mathrm{r}}\,{-}\chi$ to an intermediate frequency~$\omega_{\mathrm{if}}/2\pi\,{=}\,\SI{62.5}{\mega\hertz}$. We digitize the signals using two analog-to-digital converters (ADCs) with a sampling rate of \SI{250}{\mega\hertz} and perform digital homodyning. In addition, we can readout the resonator via a vector network analyzer (VNA) for spectroscopic analysis of the sample. From this spectroscopic analysis based on a two-tone experiment, we extract the qubit transition frequency $\omega_{\mathrm{q}}\,{=}\,\omega_{\mathrm{q},0}\sqrt{\left|\cos\left(\pi\Phi/\Phi_{0}\right)\right|}$ shown in Fig.\,\ref{fig:S02}\,(a). Here,~$\Phi_0\,{\simeq}\,\SI{2}{\femto\volt\second}$ is the magnetic flux quantum and~$\omega_{\mathrm{q},0}/2\pi\,{\simeq}\,\SI{6.92}{\giga\hertz}$ is the maximum qubit transition frequency where we perform our measurements. All experiments are carried out in the dispersive regime, where the detuning~$\delta\,{\equiv}\,\omega_{\mathrm{q}}\,{-}\,\omega_{\mathrm{r}}$ fulfills~$|\chi|\,{\ll}\,g$. Here,~\cite{Koch_2007}~$\chi\,{\equiv}\,[g^{2}/\delta][\alpha/(\delta\,{+}\,\alpha)]\,{\simeq}\,{-}2\pi\,{\times}\,\SI{3.11}{\mega\hertz}$ is the dispersive shift and $\alpha\,{\simeq}\,{-}E_{\mathrm{C}}/\hbar\,{\simeq}\,{-}2\pi\,{\times}\,\SI{315}{\mega\hertz}$ is the transmon anharmonicity. In our experiments, the average resonator population~$n_{\mathrm{r}}$ stays always below the critical photon number\,\cite{Gambetta_2006}~$n_{\mathrm{crit}}\,{=}\,\delta^{2}/4g^{2}\,{\simeq}\,40$ indicating the scale at which the dispersive approximation breaks down. We calibrate the mean photon number $n_{\mathrm{r}}$ inside the resonator by measuring the input-power-dependent AC Stark shift of the qubit frequency $\delta\omega_{\mathrm{q}}\,{=}\,2\chi n_{\mathrm{r}}$. For coherent states and shot noise, we vary the output power of the microwave source and the noise generator, respectively [see Fig.\,\ref{fig:S02}\,(b)]. For thermal states, we vary the temperature of a blackbody radiator [see Fig.\,\ref{fig:S02}\,(c)]. We provide a detailed description of noise generation in the following paragraph. Please note that we do not observe any corrections to the linear relation between qubit frequency and noise power as predicted for the strong dispersive regime~\cite{Clerk_2007}. Hence, our sample can be treated in the weak dispersive regime~$\chi/\kappa_{\mathrm{r}}\,{\ll}\,1$. \medskip

\paragraph{Generation of broadband white noise} To generate thermal states on the readout line, we use a temperature controllable \SI{30}{\decibel} attenuator integrated into the feedlines as depicted in Fig.\,\ref{fig:S01}. We can stabilize the temperature $T$ of the attenuator with an analog PID controller between \SI{0.050+-0.001}{\kelvin} and \SI{1.50+-0.01}{\kelvin}. For the coaxial cables connecting the attenuators to the sample box, we use \SI{20}{\centi\meter} of Nb/CuNi UT47, such that the sample temperature is not affected and stays stabilized at $T_{\mathrm{i}}\,{=}\,\SI{0.035+-0.001}{\kelvin}$. Thermal noise from higher temperature stages has only a negligible influence on our setup due the cryogenic attenuators and circulators in the feedlines [see Fig.\,\ref{fig:S01}]. In addition to real thermal noise radiated from the attenuators, we can add noise generated from an AFG which we upconvert to the desired noise frequency~$\omega_{\mathrm{n}}$ by mixing with a phase-modulated microwave drive. The AFG generates Gaussian shot noise with a \SI{500}{\mega\hertz} bandwidth. We additionally filter this noise before the upconversion to the carrier frequency $\omega_{\mathrm{n}}$ by two \SI{100}{\mega\hertz} low-pass filters. That way, the noise has a bandwidth of \SI{200}{\mega\hertz} and an on/off ratio of \SI{35}{\decibel}.\medskip

\paragraph{Quantum description of thermal noise} The heatable attenuator emits a voltage $V(t)\,{=}\,V_{\mathrm{vac}}[\hat{\xi}_{k}(t)\,{+}\,\hat{\xi}^{\dagger}_{k}(t)]$, which fluctuates in time with a Gaussian distribution. Here, $V_{\mathrm{vac}}$ is the vacuum amplitude of the corresponding mode. The respective field operators $\hat{\xi}^{\dagger}_{k},\,\hat{\xi}^{\phantom{\dagger}}_{k}$ create (annihilate) the individual field modes with frequencies $\omega_{k}$. Consequently, thermal noise can be described as a bosonic bath with respective Hamiltonian $\mathcal{H}_{\mathrm{bath}}\,{=}\,\sum^{\phantom{\dagger}}_{k}\hbar\omega^{\phantom{\dagger}}_{k}\hat{\xi}^{\dagger}_{k}\hat{\xi}^{\phantom{\dagger}}_{k}$. For finite temperatures, the temporal correlations of the voltage fluctuations are defined by the Hurwitz function~\cite{Kano_1962,Mehta_1963,Klein_2008}. For temperatures in the kelvin range, the  Hurwitz function yields a sub-nanosecond, i.e.~negligible, coherence time of thermal fields, which can therefore be described by a $\delta$-function $\hat{\xi}^{\phantom{\dagger}}_{k}(t)\hat{\xi}^{\dagger}_{k}(t^{\prime})\,{=}\,\delta(t\,{-}\,t^{\prime})$.

\section{Correlators for shot-noise, thermal, and coherent states}

To derive the photon number correlator $\mathcal{C}(\tau)$ inside the resonator for shot-noise, thermal, and coherent states, we use input-output theory for a single mode $\hat{a}$ of the resonator described by the Hamiltonian $\mathcal{H}_{\mathrm{r}}\,{=}\,\hbar\omega_{\mathrm{r}}\hat{a}^{\dagger}\hat{a}$.\medskip

\paragraph{Correlator for incoherently driven resonators} We start with deriving the correlator for broadband white noise, which could be thermal (i.e.~super-Poissonian) or shot noise (i.e.~Poissonian). The noise field occupies incoming modes $\hat{b}_{\mathrm{in}}(\omega,t)$ of an open transmission line, which is coupled to the resonator via a coupling capacitor with strength $\tilde{g}(\omega)$. Using the resonator density of states $D(\omega)$, the amplitude-damping of the resonator is given as $\kappa_{\mathrm{x}}/2\,{=}\,\pi D(\omega_{\mathrm{r}})|\tilde{g}(\omega_{\mathrm{r}})|^{2}$. The transmission line modes describe a force~\cite{Devoret_2014} $F(t)\,{=}\,\imath\sum_{\omega}\tilde{g}(\omega)\hat{b}_{\mathrm{in}}(\omega,0)e^{{-}\imath\omega t}$ acting on the resonator such that the equation of motion becomes~\cite{Bertet_2005} $\partial\hat{a}/\partial t\,{=}\,{-}[\imath\omega_{\mathrm{r}}\,{+}\,\kappa_{\mathrm{x}}]\hat{a}(t)/2\,{-}\,F(t)$. In the Markov approximation, this equation solves in the Heisenberg picture to
\begin{equation}
 \hat{a}(\tau) = e^{-(\imath\omega_{\mathrm{r}} + \kappa_{\mathrm{x}}/2)\tau}\left[\hat{a}(0) - \int_{0}^{\tau}\mathrm{d}t~e^{-(\imath\omega_{\mathrm{r}} + \kappa_{\mathrm{x}}/2)\tau} F(t)\right]\,.
\label{eqn:ab}
\end{equation}
Using this equation, we calculate the photon-photon time correlator
\begin{align}
 \mathcal{C}(\tau) &\equiv \langle\delta n_{\mathrm{r}}(0)\delta n_{\mathrm{r}}(\tau)\rangle = \langle\hat{a}^{\dagger}(0)\hat{a}(0)\hat{a}^{\dagger}(\tau)\hat{a}(\tau)\rangle - \langle\hat{a}^{\dagger}(0)\hat{a}(0)\rangle^{2} \notag\\
& =  \langle\hat{a}^{\dagger}(0)\hat{a}(0)\hat{a}^{\dagger}(0)\hat{a}(0)\rangle e^{-\kappa_{\mathrm{x}}\tau} - \langle\hat{a}^{\dagger}\hat{a}\rangle^{2}\notag\\
  &-e^{-\kappa_{\mathrm{x}}\tau}\int_{0}^{\tau}\mathrm{d}t~\langle\hat{a}^{\dagger}(0)\hat{a}(0)F^{\dagger}(t)\hat{a}(0) \rangle e^{(-\imath\omega_{\mathrm{r}} + \kappa_{\mathrm{x}}/2)t}\hspace{2.95cm}(i)\notag\\
  &+e^{-\kappa_{\mathrm{x}}\tau}\int_{0}^{\tau}\mathrm{d}t~\langle\hat{a}^{\dagger}(0)\hat{a}(0)\hat{a}^{\dagger}(0)F(t) \rangle e^{(\imath\omega_{\mathrm{r}} + \kappa_{\mathrm{x}}/2)t}\hspace{3.17cm}(ii)\notag\\
  &+e^{-\kappa_{\mathrm{x}}\tau}\int_{0}^{\tau}\int_{0}^{\tau}\mathrm{d}t\mathrm{d}t^{\prime}~\langle\hat{a}^{\dagger}(0)\hat{a}(0)F^{\dagger}(t)F(t^{\prime}) \rangle e^{-\imath\omega_{\mathrm{r}}(t - t^{\prime})} e^{\kappa_{\mathrm{x}}/2(t + t^{\prime})}\hspace{0.7cm}(iii)\,.
\label{eqn:4thmom}
\end{align}
Since external force and resonator are uncorrelated at $t\,{=}\,0$, we write the two terms $(i)$ and $(ii)$ in Eq.\,(\ref{eqn:4thmom}) as~\cite{Bertet_2005}
\begin{align}
  (i) &= -\imath\sum_{\omega}\tilde{g}(\omega)e^{-\kappa_{\mathrm{x}}\tau}\int_{0}^{\tau}\mathrm{d}t~\langle\hat{a}^{\dagger}(0)\hat{a}(0)\hat{a}(0)\rangle~\langle \hat{b}_{\mathrm{in}}^{\dagger}(\omega,0)\rangle e^{\imath\omega t} e^{(-\imath\omega_{\mathrm{r}} + \kappa_{\mathrm{x}}/2)t}\notag\\
  (ii) & = +\imath\sum_{\omega}\tilde{g}(\omega)e^{-\kappa_{\mathrm{x}}\tau}\int_{0}^{\tau}\mathrm{d}t~\langle\hat{a}^{\dagger}(0)\hat{a}(0)\hat{a}^{\dagger}(0)\rangle~\langle \hat{b}_{\mathrm{in}}(\omega,0)\rangle e^{{-}\imath\omega t}e^{(\imath\omega_{\mathrm{r}} + \kappa_{\mathrm{x}}/2)t}\notag\,.
\end{align}
Here, one immediately sees that these expressions become zero for uncorrelated signals with zero mean (noise), characterized by $\langle \hat{b}_{\mathrm{in}}^{\dagger}\rangle\,{=}\,0\,{=}\,\langle \hat{b}_{\mathrm{in}}\rangle$. Hence, the relevant part for $\mathcal{C}(\tau)$ is determined by the last part $(iii)$ in Eq.\,(\ref{eqn:4thmom}). Assuming that both thermal and shot noise have a white frequency distribution over the resonator bandwidth, i.e., $\langle F^{\dagger}(t)F(t^{\prime})\rangle\,{=}\,\kappa_{\mathrm{x}}\langle\hat{b}_{\mathrm{in}}^{\dagger}(\omega,0)\hat{b}_{\mathrm{in}}(\omega,0)\rangle\delta(t\,{-}\,t^{\prime})$, on resonance, part $(iii)$ simplifies to
\begin{align}
 (iii) &= \kappa_{\mathrm{x}}\langle\hat{b}_{\mathrm{in}}^{\dagger}(\omega_{\mathrm{r}},0)\hat{b}_{\mathrm{in}}(\omega_{\mathrm{r}},0)\rangle\langle\hat{a}^{\dagger}(0)\hat{a}(0)\rangle e^{-\kappa_{\mathrm{x}}\tau}\int_{0}^{\tau}\int_{0}^{\tau}\mathrm{d}t\mathrm{d}t^{\prime}~\delta(t\,{-}\,t^{\prime}) e^{-\imath\omega_{\mathrm{r}}(t - t^{\prime})} e^{\kappa_{\mathrm{x}}(t + t^{\prime})/2}\notag\\
  &= \kappa_{\mathrm{x}}\langle\hat{b}_{\mathrm{in}}^{\dagger}(\omega_{\mathrm{r}},0)\hat{b}_{\mathrm{in}}(\omega_{\mathrm{r}},0)\rangle\langle\hat{a}^{\dagger}(0)\hat{a}(0)\rangle e^{-\kappa_{\mathrm{x}}\tau}\int_{0}^{\tau}\mathrm{d}t~e^{\kappa_{\mathrm{x}}t}\notag\\
  &= ~~~\langle\hat{b}_{\mathrm{in}}^{\dagger}(\omega_{\mathrm{r}},0)\hat{b}_{\mathrm{in}}(\omega_{\mathrm{r}},0)\rangle\langle\hat{a}^{\dagger}(0)\hat{a}(0)\rangle [1 - e^{-\kappa_{\mathrm{x}}\tau}]\,.
\end{align}
Using this expression, and equilibrium states, $\langle\hat{b}_{\mathrm{in}}^{\dagger}(\omega_{\mathrm{r}},0)\hat{b}_{\mathrm{in}}(\omega_{\mathrm{r}},0)\rangle\,{=}\,\langle\hat{a}^{\dagger}\hat{a}\rangle\,{\equiv}\,n_{\mathrm{r}}$, Eq.\,(\ref{eqn:4thmom}) simplifies to $\mathcal{C}(\tau)\,{=}\,[\langle\hat{a}^{\dagger}\hat{a}\hat{a}^{\dagger}\hat{a}\rangle\,{-}\,\langle\hat{a}^{\dagger}\hat{a}\rangle^{2}]\exp({-}\kappa_{\mathrm{x}}\tau)\,{=}\,\mathrm{Var}(n_{\mathrm{r}})\exp({-}\kappa_{\mathrm{x}}\tau)$. Hence, the correlator for white noise is given as the photon number variance which decays at the energy decay rate $\kappa_{\mathrm{x}}$. Because we did not make any assumptions on the distribution of the noise (Poissonian, super-Poissonian, etc.), this statement holds for shot noise with Var$(n_{\mathrm{r}})\,{=}\,n_{\mathrm{r}}$ and  $\mathcal{C}^{\mathrm{sh}}(\tau)\,{=}\,n_{\mathrm{r}}\exp({-}\kappa_{\mathrm{x}}\tau)$ as well as for thermal noise with Var$(n_{\mathrm{r}})\,{=}\,n_{\mathrm{r}}^{2}\,{+}\,n_{\mathrm{r}}$ and  $\mathcal{C}^{\mathrm{th}}(\tau)\,{=}\,(n_{\mathrm{r}}^{2}\,{+}\,n_{\mathrm{r}})\exp({-}\kappa_{\mathrm{x}}\tau)$. We discuss the effect of the broadband nature of noise fields entering the resonator environment modelled by $D(\omega)$ in the next section.\medskip

\paragraph{Correlator for coherently driven resonators} For coherently driven resonators, it is convenient to describe the transmission line modes as $\hat{b}_{\mathrm{in}}(\omega,t)\,{=}\,e^{\imath\omega t}[\bar{b}_{\mathrm{in}}\,{+}\,\hat{\xi}(t)]$. Here, $\bar{b}_{\mathrm{in}}$ describes a classical coherent drive with frequency $\omega\,{=}\,\omega_{\mathrm{r}}\,{-}\,\delta_{\mathrm{r}}$. For a coherent input field, we assume that $\hat{\xi}^{(\dagger)}(t)$ are just vacuum fluctuations~\cite{Blais_2004,Clerk_2010,Devoret_2014}. In a frame rotating at $\omega$, the field inside the cavity is given as $\hat{a}(\tau)\,{=}\,\bar{a}\,{+}\,\hat{d}(\tau)$, where~\cite{Clerk_2010}
\begin{align}
 \bar{a} &= \frac{-\sqrt{\kappa}}{\imath\delta_{\mathrm{r}} + \kappa/2}\bar{b}_{\mathrm{in}}\,,\\
\hat{d}(\tau) &= -\sqrt{\kappa}\int_{-\infty}^{\tau}\mathrm{d}t\,e^{-(\imath\delta_{\mathrm{r}}+\kappa/2)(\tau - t)}\hat{\xi}(t)\,.
\end{align}
We can use this expression to calculate the correlator $\mathcal{C}^{\mathrm{coh}}(\tau)\,{=}\,n_{\mathrm{r}}\langle\hat{d}(0)\hat{d}^{\dagger}(\tau)\rangle$, which decays at the amplitude decay rate $\kappa_{\mathrm{x}}/2$ since~\cite{Blais_2004,Clerk_2010} $\langle\hat{d}^{\dagger}(0)\hat{d}(t)\rangle\,{=}\,\exp({-}\imath\delta_{\mathrm{r}}t\,{-}\,\kappa_{\mathrm{x}}|\tau|/2)$. Hence, on resonance, the correlator for a coherent state $\mathcal{C}^{\mathrm{coh}}(\tau)\,{=}\,n_{\mathrm{r}}\exp({-}\kappa_{\mathrm{x}}|\tau|/2)$ decays twice as slow as the correlator for white noise.

\section{Photon-number-dependent qubit dephasing rate}

To derive the photon-number-dependent dephasing rate in the dispersive regime, we start with the system Hamiltonian $\mathcal{H}_{\mathrm{tot}}\,{=}\,\mathcal{H}_{\mathrm{JC}}\,{+}\,\mathcal{H}_{\mathrm{d}}$ comprising the Jaynes-Cummings Hamiltonian $\mathcal{H}_{\mathrm{JC}}\,{=}\,\mathcal{H}_{\mathrm{r}}\,{+}\,\mathcal{H}_{\mathrm{q}}\,{+}\,\mathcal{H}_{\mathrm{int}}$ and a driving part $\mathcal{H}_{\mathrm{d}}$. Here, $\mathcal{H}_{\mathrm{q}}\,{=}\,\hbar\omega_{\mathrm{q}}\hat{\sigma}_{\mathrm{z}}/2$ is the bare qubit Hamiltonian, $\mathcal{H}_{\mathrm{int}}\,{=}\,{-}\hbar g(\hat{a}^{\dagger}\hat{\sigma}_{-}\,{+}\,\hat{a}\hat{\sigma}_{+})$ describes the qubit-resonator interaction using the Pauli operators $\hat{\sigma}_{i}$ and the resonator mode $\hat{a}$ with mean occupation $n_{\mathrm{r}}\,{=}\,\langle\hat{a}^{\dagger}\hat{a}\rangle$. The Hamiltonian for external drives with amplitudes $\varepsilon_{j}(t)$ reads $\mathcal{H}_{\mathrm{d}}\,{=}\,\sum_{j}\hbar\varepsilon_{j}(t)(\hat{a}^{\dagger}e^{{-}\imath\omega_{j}t}\,{+}\,\hat{a}e^{{+}\imath\omega_{j}t})$. We use two different coherent drives, one to read out the resonator $(j\,{=}\,\mathrm{r})$ and one to drive the qubit  $(j\,{=}\,\mathrm{d})$. Furthermore, the noise field $\hat{b}_{\mathrm{in}}(\omega)$ can be treated as an incoherent external drive $(j\,{=}\,\mathrm{in})$ such that $\varepsilon_{\mathrm{in}}(t)e^{{-}\imath\omega_{j}t}\,{\mapsto}\,\pi^{-1}\kappa_{\mathrm{x}}e^{{-}\imath\omega t}\int\mathrm{d}\omega\,D(\omega)\hat{b}^{\dagger}_{\mathrm{in}}(\omega,t)\hat{b}^{\phantom{\dagger}}_{\mathrm{in}}(\omega,t)$. We assume that this drive is weak and $\delta$-correlated in time, i.e., $\langle \varepsilon_{\mathrm{in}}^{\star}(t)\varepsilon_{\mathrm{in}}^{\phantom{\star}}(t^{\prime})\rangle\,{\propto}\,\delta(t\,{-}\,t^{\prime})$. The dynamics of the qubit-resonator system can conveniently be described using the master equation\,\cite{Gambetta_2006,Gambetta_2008,Boissonneault_2009}
\begin{equation}
\hbar\partial\hat{\rho}/\partial t = -\imath[\mathcal{H}_{\mathrm{tot}},\hat{\rho}] + \kappa_{\mathrm{x}}\mathcal{D}(\hat{a})\hat{\rho} + \gamma_{1}^{\mathrm{tot}}\mathcal{D}(\hat{\sigma}_{-})\hat{\rho} + \gamma_{\phi}^{\mathrm{tot}}\mathcal{D}(\hat{\sigma}_{\mathrm{z}})\hat{\rho}/2\,.
\label{eqn:denstot}
\end{equation}
Here,~$\hat{\rho}\,{=}\,\mathrm{Tr}(\hat{\rho})$ is the system density matrix and~$\mathcal{D}(\hat{L})\,{=}\,[2\hat{L}\hat{\rho}\hat{L}^{\dagger}\,{-}\,\hat{L}^{\dagger}\hat{L}\hat{\rho}\,{-}\,\hat{\rho}\hat{L}^{\dagger}\hat{L}]/2$ is the Lindblad damping operator~\cite{Lindblad_1976}, which describes effects of the bath in the Markov approximation. The qubit is characterized by the total energy decay rate $\gamma_{1}^{\mathrm{tot}}$ and the total dephasing rate $\gamma_{\phi}^{\mathrm{tot}}$. To study the effect of photon number fluctuations, we transform $\mathcal{H}_{\mathrm{tot}}$ into the dispersive regime, which yields
\begin{equation}
 \mathcal{H}_{\mathrm{eff}} = \mathcal{H}_{\mathrm{r}} + \mathcal{H}_{\mathrm{q}} + \frac{\hbar}{2}(\chi + 2\chi\hat{a}^{\dagger}\hat{a})\hat{\sigma}_{\mathrm{z}} + \sum_{j}\hbar\varepsilon_{j}(t)(\hat{a}^{\dagger}e^{{-}\imath\omega_{j}t} + \hat{a}e^{{+}\imath\omega_{j}t}) + \sum_{j}\frac{\hbar\varepsilon_{j}(t)}{\delta}(\hat{\sigma}_{+}e^{{-}\imath\omega_{j}t} + \hat{\sigma}_{-}e^{{+}\imath\omega_{j}t})\,.
\end{equation}

After this transformation and tracing out the resonator yields the laboratory frame master equation for the qubit~\cite{Gambetta_2008}
\begin{equation}
\frac{\partial\hat{\rho}_{\mathrm{q}}^{\mathrm{D}}}{\partial t} = \frac{1}{\imath\hbar}[\mathcal{H}_{\mathrm{eff}},\hat{\rho}_{\mathrm{q}}^{\mathrm{D}}] + \gamma_{1}^{\mathrm{tot}}\mathcal{D}(\hat{\sigma}_{-}) + \frac{\gamma_{\phi}^{\mathrm{tot}}}{2}\mathcal{D}(\hat{\sigma}_{\mathrm{z}})\,,
\label{eqn:densqub}
\end{equation}
where $\hat{\rho}_{\mathrm{q}}^{\mathrm{D}}$ describes the qubit in the dispersive regime. In this regime, the total qubit decay rate $\gamma_{1}^{\mathrm{tot}}$ and dephasing rate $\gamma_{\phi}^{\mathrm{tot}}$ are dependent on the photon number $n_{\mathrm{r}}$ in the resonator~\cite{Boissonneault_2009}. While the change of the qubit decay rate is due to dressing of states and due to frequency components at $\omega_{\mathrm{q}}$, the change of the dephasing rate is due to photon number fluctuations characterized by $\mathcal{C}(\tau)$. In Fig.\,\ref{fig:S02}\,(d) we show the linear dependence between the qubit decay rate $\gamma_{1}^{\mathrm{tot}}(n_{\mathrm{r}})\,{=}\,\gamma_{1}\,{+}\,\gamma_{1}^{\mathrm{d}}n_{\mathrm{r}}$ and the emitter temperature $T$. As discussed in detail in Ref.~\citealp{Goetz_2016b}, we find the intrinsic decay rate $\gamma_{1}/2\pi\,{\simeq}\,\SI{3.9}{\mega\hertz}$ and the additional decay per thermal photon $\gamma_{1}^{\mathrm{d}}/2\pi\,{\simeq}\,\SI{800}{\kilo\hertz}$. For coherent states and shot noise, we find~\cite{Goetz_2016b} $\gamma_{1}^{\mathrm{d}}/2\pi\,{\simeq}\,\SI{-30}{\kilo\hertz}$. Please note that $\gamma_{1}$ strongly exceeds the Purcell decay rate $\gamma_{\mathrm{P}}\,{=}\,\kappa_{\mathrm{tot}}g^{2}/\delta^{2}\,{\simeq}\,2\pi\,{\times}\,\SI{53}{\kilo\hertz}$. The qubit dephasing rate $\gamma_{\phi}^{\mathrm{tot}}\,{=}\,\gamma_{\varphi0}\,{+}\,\gamma_{\varphi n}(n_{\mathrm{r}})$ comprises the bare qubit dephasing plus the power broadening $\gamma_{\varphi n}(n_{\mathrm{r}})$. For a small cavity pull $(|\chi/\kappa_{\mathrm{x}}|\,{\ll}\,1)$, we can make a Gaussian approximation for the dephasing rate $\gamma_{\varphi n}(n_{\mathrm{r}})$ as discussed in the following paragraph. For our sample, we find $|\chi/\kappa_{\mathrm{x}}|\,{\simeq}\,0.35$, which leads to small corrections because the effective resonator frequency is different if the qubit is in the ground or the excited state. We discuss the corresponding corrections for broadband fields in the last paragraph of this section.\medskip

\begin{figure*}[b]
\includegraphics{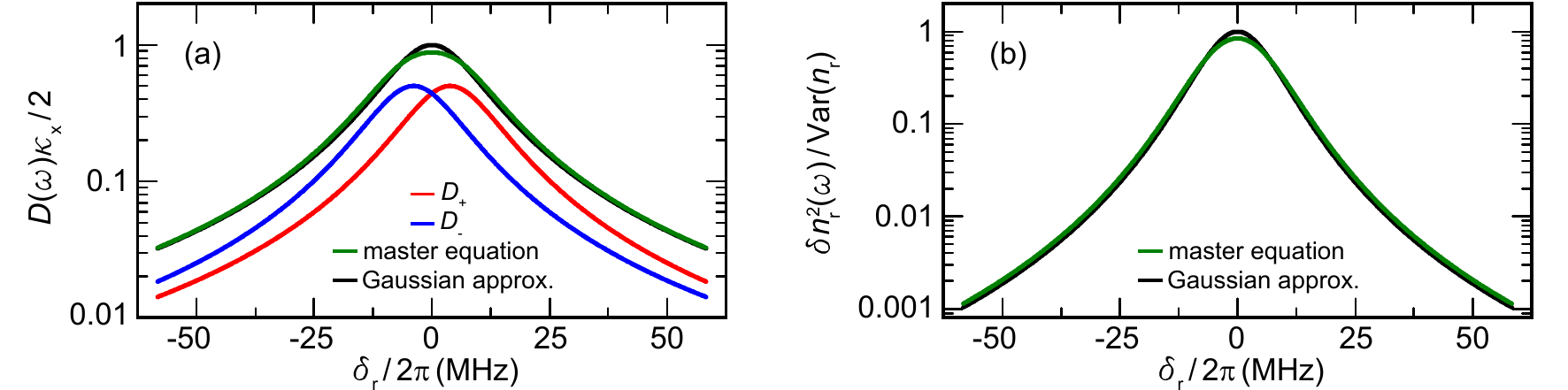}
\caption{\label{fig:S03} (a) Normalized resonator density of states $D(\omega)\kappa_{\mathrm{x}}/2$. For the Gaussian approximation, we use $D(\omega)\,{=}\,\mathcal{F}(\omega)$. For $D_{\pm}(\omega)$, we plot Eq.\,(\ref{eqn:D+-})/2, because we calibrate the photon number in a steady state protocol $\langle\hat{\sigma}_{\mathrm{z}}\rangle\,{=}\,0.5$. For the master equation, we plot $D(\omega)\,{=}\,[D_{+}(\omega)\,{+}\,D_{-}(\omega)]/2$. (b) Photon fluctuations $\delta n_{\mathrm{r}}^{2}(\omega)$ calculated for the parameters used in panel (a).}
\end{figure*}

\paragraph{Qubit dephasing under the Gaussian approximation} If we assume the resonator pull on the qubit to be weak $(|\chi/\kappa_{\mathrm{x}}|\,{\ll}\,1)$, we can assume that after a time $\tau$, the random phase accumulated~\cite{Blais_2004} $\delta\varphi(\tau)\,{=}\,2\chi\int_{0}^{\tau}\mathrm{d}t\,\delta n(t)$ is Gaussian distributed. In this case, the cumulant expansion is exact and one obtains~\cite{Gambetta_2006}
\begin{equation}
 \langle\hat{\sigma}_{-}(\tau)\hat{\sigma}_{+}(0)\rangle = \exp\left[-\gamma_{2}\tau-\frac{\langle\delta\varphi^{2}\rangle}{2}\right] = \exp\left[-\gamma_{2}\tau-2\chi^{2}\int_{0}^{\tau}\mathrm{d}t\,\mathcal{C}(t)\right]\,,
\label{eqn:s+s-}
\end{equation}
where $\gamma_{2}\,{=}\,\gamma_{1}^{\mathrm{tot}}/2\,{+}\,\gamma_{\varphi0}$ describes the qubit decoherence. Equation\,(\ref{eqn:s+s-}) leads to the qubit dephasing rates defined in Eqs.\,(1)\,{-}\,(3) in the article. To calibrate the mean photon number $n_{\mathrm{r}}$, we measure the photon number dependent AC Stark shift of the qubit frequency as a function of noise power in a steady state configuration. The externally applied broadband input field $\hat{b}_{\mathrm{in}}$ is linked via Eq.\,(\ref{eqn:ab}), to the intra-resonator mode $\hat{a}$. Because we measure in a steady state, $\langle\hat{b}_{\mathrm{in}}\rangle\,{=}\,0$ and $\langle\hat{b}^{\dagger}_{\mathrm{in}}\hat{b}^{\phantom{\dagger}}_{\mathrm{in}}\rangle$ has a constant mean. Further, because we assume white noise, $\hat{b}_{\mathrm{in}}$ also has no frequency dependence in the relevant frequency regime. Within these limits and for negligible internal resonator losses, the broadband noise covering the resonator density of states $D(\omega)$ yields~\cite{Goetz_2016b}
\begin{equation}
 \delta\omega_{\mathrm{q}} = 2\chi\hat{\sigma}_{\mathrm{z}} \frac{\langle\hat{b}^{\dagger}_{\mathrm{in}}\hat{b}^{\phantom{\dagger}}_{\mathrm{in}}\rangle}{\pi}\int\mathrm{d}\omega\,D(\omega) = 2\chi\hat{\sigma}_{\mathrm{z}}\langle\hat{a}^{\dagger}\hat{a}\rangle\, \equiv 2\chi\hat{\sigma}_{\mathrm{z}}n_{\mathrm{r}}\,.
\label{eqn:n}
\end{equation}
The above calibration method can also be obtained using a Wigner function approach~\cite{Clerk_2007} for the qubit off-diagonal elements in the small pull limit $|\chi/\kappa_{\mathrm{x}}|\,{\ll}\,1$. For larger pulls, this approach predicts a deviation from the linear trend predicted by Eq.\,(\ref{eqn:n}). However, since we do not observe any non-linear trend in our Stark shift measurements [cf.~Fig.\,\ref{fig:S02}\,(b) and (c)], we conclude that our sample can be still treated in the small pull limit. Within this approximation, we do not separate between the two cases when the qubit is in the ground or the excited state such that $D(\omega)$ is simply given by the Lorentzian filter function $\mathcal{F}(\omega)\,{=}\,(\kappa_{\mathrm{x}}/2)/[(\kappa_{\mathrm{x}}/2)^{2}\,{+}\,\delta_{\mathrm{r}}^{2}]$, where $\delta_{\mathrm{r}}\,{=}\,\omega_{\mathrm{r}}\,{-}\,\omega$ is the detuning to the resonator frequency [see Fig.\,\ref{fig:S03}\,(a)]. The broadband noise induces dephasing relative to photon number fluctuations $\delta n_{\mathrm{r}}^{2}(\omega)\,{=}\,(\kappa_{\mathrm{x}}/2)^{2}\mathcal{F}(\omega)^{2}[\langle\hat{b}^{\dagger}_{\mathrm{in}}\hat{b}^{\phantom{\dagger}}_{\mathrm{in}}\hat{b}^{\dagger}_{\mathrm{in}}\hat{b}^{\phantom{\dagger}}_{\mathrm{in}}\rangle\,{-}\,\langle\hat{b}^{\dagger}_{\mathrm{in}}\hat{b}^{\phantom{\dagger}}_{\mathrm{in}}\rangle^{2}]$ shown in Fig.\,\ref{fig:S03}\,(b). Hence, the effective dephasing rate reads
\begin{equation}
 \gamma_{\varphi n}^{\mathrm{eff}} = \theta_{0}^{2}\frac{4}{\pi}\int\mathrm{d}\omega\,\delta n_{\mathrm{r}}^{2}(\omega) = \mathrm{Var}(n_{\mathrm{r}})\kappa_{\mathrm{x}}\theta_{0}^{2}\,.
\label{eqn:Gauss}
\end{equation}
Because the assumption of a Lorentzian line shape $D(\omega)\,{=}\,\mathcal{F}(\omega)$ is only approximately true for our experimental parameters $(|\chi/\kappa_{\mathrm{x}}|\,{\simeq}\,0.35\,{\lesssim}\,1)$, we evaluate corrections to Eq.\,(\ref{eqn:Gauss}) in the following paragraph. These corrections are due to the fact that the resonator has a different frequency when the qubit is in its ground or excited state.\medskip

\paragraph{Qubit dephasing using the full master equation} For increasing cavity pull $|\chi/\kappa_{\mathrm{x}}|$, the effective resonator frequency is different if the qubit is in its ground or excited state. We account for this circumstance using the steady state fields~\cite{Gambetta_2006} $\langle\hat{a}_{\pm}(\omega)\rangle\,{=}\,{-}\imath\langle\hat{b}_{\mathrm{in}}(\omega)\rangle\kappa_{\mathrm{x}}/(\kappa_{\mathrm{x}}\,{\pm}\,\imath2\chi\,{+}\,2\delta_{\mathrm{r}})$ if the qubit is in the excited $(+)$ or ground $(-)$ state, respectively. The two situations can be modelled by the resonator density of states
\begin{equation}
 D_{\pm}(\omega) = \frac{\kappa_{\mathrm{x}}/2}{\kappa_{\mathrm{x}}^{2}/4 + (\delta_{\mathrm{r}} \pm \chi)^{2}}\,
\label{eqn:D+-}
\end{equation}
shown in Fig.\,\ref{fig:S03}\,(a). From $D_{\pm}(\omega)$, we calculate the mean photon numbers $n_{+}$ and $n_{-}$ via Eq.\,(\ref{eqn:n}) and calibrate the effective resonator occupation $n_{\mathrm{cal}}$ as follows. Because we use a steady-state drive when calibrating the photon number, the qubit is in an equal superposition state leading to $n_{\mathrm{cal}}\,{=}\,(n_{+}\,{+}\,n_{-})/2$. In this case and for constant noise power $\langle\hat{b}^{\dagger}_{\mathrm{in}}\hat{b}^{\phantom{\dagger}}_{\mathrm{in}}\rangle$, we obtain $n_{\mathrm{cal}}\,{\approx}\,n_{\mathrm{r}}$ to a very good approximation as indicated by the black and the green lines in Fig.\,\ref{fig:S03}\,(a). Accounting for the frequency dependence of $\hat{a}_{\pm}$, we find for the photon number fluctuations [see Fig.\,\ref{fig:S03}\,(b)]
\begin{equation}
  \delta n_{\mathrm{r}}^{2}(\omega) = \frac{\kappa_{\mathrm{x}}^{2}}{4}\frac{[D_{+}(\omega) + D_{-}(\omega)]}{\kappa_{\mathrm{x}}^{2}/4 + \delta_{\mathrm{r}}^{2} + \chi^{2}}  [\langle\hat{b}^{\dagger}_{\mathrm{in}}\hat{b}^{\phantom{\dagger}}_{\mathrm{in}}\hat{b}^{\dagger}_{\mathrm{in}}\hat{b}^{\phantom{\dagger}}_{\mathrm{in}}\rangle - \langle\hat{b}^{\dagger}_{\mathrm{in}}\hat{b}^{\phantom{\dagger}}_{\mathrm{in}}\rangle^{2}]\,.
\end{equation}
With this expression, we calculate the dephasing rate $\gamma_{\varphi n}^{\mathrm{master}}\,{=}\,(2|\chi/\kappa_{\mathrm{x}}|)^{2}(\pi/4)\int\mathrm{d}\omega\,\delta n_{\mathrm{r}}^{2}(\omega)$. For the experimtal parameters stated above, we find the relative error $(\gamma_{\varphi n}^{\mathrm{eff}}\,{-}\,\gamma_{\varphi n}^{\mathrm{master}})/\gamma_{\varphi n}^{\mathrm{eff}}\,{\simeq}\,0.04$. Hence, the Gaussian approximation made in the article is well justified but would explain a decrease of the actually measured dephasing rate rather than an increase.

\section{Correlators for attenuated and amplified microwave fields}

\begin{figure}[b]
\includegraphics{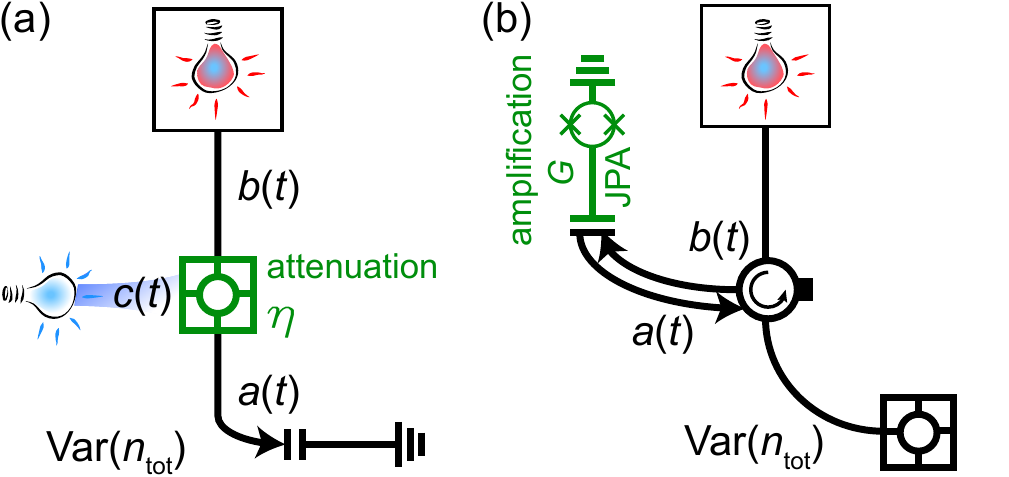}
\caption{\label{fig:S04} (a) Beam splitter model to calculate the variance of attenuated thermal and coherent fields. (b) Input-output model to calculate the variance of amplified thermal fields.}
\end{figure}

In this section, we calculate the variance of the attenuated thermal states based on a beam splitter model depicted in Fig.\,\ref{fig:S04}\,(a). Additionally, we calculate the variance of an amplified thermal fields [see Fig.\,\ref{fig:S04}\,(b)], which is relevant for reconstruction setups~\cite{Menzel_2012,Fedorov_2016} using parametric amplifiers as preamplifiers.\\
To calculate the photon number variance of an attenuated thermal field, we use the beam splitter model depicted in Fig.\,\ref{fig:S04}\,(a). Here, we assume a thermal state generated at a higher temperature stage, which is subsequently attenuated inside the cryostat with attenuation $\eta\,{\leq}\,1$. We describe the thermal state emitted from the temperature controllable attenuator with the bosonic operators $\hat{b}^{\dagger}$ and $\hat{b}$ and model the parasitic thermal influence of the cryostat as a beam splitter which mixes a weak thermal state described by $\hat{c}^{\dagger}$ and $\hat{c}$ to the incoming state. That way, we obtain the mixed state described with the operator $\hat{a}(t)\,{=}\,\sqrt{\eta}\hat{b}(t)\,{+}\,\sqrt{1\,{-}\,\eta}\hat{c}(t)$. We define the photon numbers $n_{\mathrm{b}}\,{=}\,\langle\hat{b}^{\dagger}\hat{b}\rangle$, $n_{\mathrm{n}}\,{=}\,\langle\hat{c}^{\dagger}\hat{c}\rangle$ and the total photon number $n_{\mathrm{tot}}\,{=}\,\langle\hat{a}^{\dagger}\hat{a}\rangle\,{=}\,B^{2}n_{\mathrm{b}}\,{+}\,C^{2}n_{\mathrm{n}}$. Then, following Ref.~\onlinecite{Blais_2004}, the photon number variance is defined by the correlator
\begin{align}
\mathcal{C}^{\mathrm{th}}(\tau)&= \langle(\hat{a}^{\dagger}\hat{a} - \langle \hat{a}^{\dagger}\hat{a}\rangle)^{2} \rangle\exp(-\kappa_{\mathrm{x}}\tau)\notag\\
&=[ \eta^{2}n_{\mathrm{b}}^{2}+\eta n_{\mathrm{b}} + 2\eta(1-\eta)n_{\mathrm{b}} n_{\mathrm{n}} + (1-\eta)^{2}n_{\mathrm{n}}^{2}+(1-\eta)n_{\mathrm{n}}]\exp(-\kappa_{\mathrm{x}}\tau)\,.
\label{eqn:Cth}
\end{align}
From this equation we see that the beam splitter model predicts the thermal photon statistics of the emitted field for $\eta\,{\mapsto}\,1$ (no background field) and the thermal photon statistics of the background field for $\eta\,{\mapsto}\,0$ (strong  background field). In a similar way, we calculate the correlator of an attenuated coherent field
\begin{align}
 \mathcal{C}^{\mathrm{coh}}(\tau) &= \langle(\hat{a}^{\dagger}\hat{a} - \langle \hat{a}^{\dagger}\hat{a}\rangle)^{2} \rangle\exp(-\kappa_{\mathrm{x}}\tau/2)\notag\\
&= [\eta n_{\mathrm{b}} + 2\eta(1-\eta)n_{\mathrm{b}} n_{\mathrm{n}} +(1-\eta)^{2}n_{\mathrm{n}}^{2}+(1-\eta)n_{\mathrm{n}}]\exp(-\kappa_{\mathrm{x}}\tau/2)\,,
\end{align}
which approaches the variance of a coherent state for $\eta\,{\mapsto}\,1$ and the thermal photon statistics of the cold attenuator for $\eta\,{\mapsto}\,0$.

Similar to the calculations of attenuated propagating microwaves, we calculate the variance of an amplified thermal field using input-output relations. Following Ref.~\onlinecite{Candia_2014} and Ref.~\onlinecite{Caves_1982}, we describe the amplified field by the operator $\hat{a}(t)\,{=}\,\sqrt{G}\hat{b}\,{+}\,\sqrt{G-1}\hat{c}^{\dagger}$, where $n_{\mathrm{bs}}\,{=}\,\langle\hat{a}^{\dagger}\hat{a}\rangle$. The quantity $n_{\mathrm{n}}\,{=}\,\langle\hat{c}^{\dagger}\hat{c}\rangle$ describes the noise photons added by the JPA, which we assume to be thermal. Based on these assumptions, we obtain the photon number variance
\begin{align}
 \mathrm{Var}(n_{\mathrm{bs}}) &= \langle(\hat{a}^{\dagger}\hat{a} - \langle \hat{a}^{\dagger}\hat{a}\rangle)^{2} \rangle\notag\\
&= G^{2}n_{\mathrm{jpa}}^{2} + G^{2}n_{\mathrm{jpa}} + G(G-1)n_{\mathrm{jpa}}  +  2G(G-1)n_{\mathrm{jpa}}n_{\mathrm{n}}\notag\\
&+(G-1)^{2}n_{\mathrm{n}}^{2} + (G-1)^{2}n_{\mathrm{n}} + G(G-1)n_{\mathrm{n}} + G(G-1)\,,
\label{eqn:Cjpa}
\end{align}
which approaches the variance $n_{\mathrm{jpa}}^{2}\,{+}\,n_{\mathrm{jpa}}$ of a thermal state for $G\,{\mapsto}\,1$ (no amplification). For strong amplification $(G\,{\gg}\,1)$, we obtain $\mathrm{Var}(n_{\mathrm{bs}})\,{\approx}\,G^{2}(n_{\mathrm{jpa}}\,{+}\,n_{\mathrm{n}}\,{+}\,1)^{2}\,{=}\,n_{\mathrm{bs}}^{2}$. As a consequence, the unnormalized $g^{(2)}$ function of the amplified field becomes
\begin{equation}
\tilde{g}^{(2)}(0) \equiv n_{\mathrm{bs}}^{2}g^{(2)}(0) = \mathrm{Var}(n_{\mathrm{bs}}^{})\,{-}\,n_{\mathrm{bs}}^{}\,{+}\,n_{\mathrm{bs}}^{2} \approx 2G^{2}(n_{\mathrm{jpa}}\,{+}\,n_{\mathrm{n}}\,{+}\,1)^{2}\,.
\end{equation}
We fit this relation to our data as discussed in the article.

\section{Experimental setup and characterization of the JPA sample}

\begin{figure*}[b]
\includegraphics{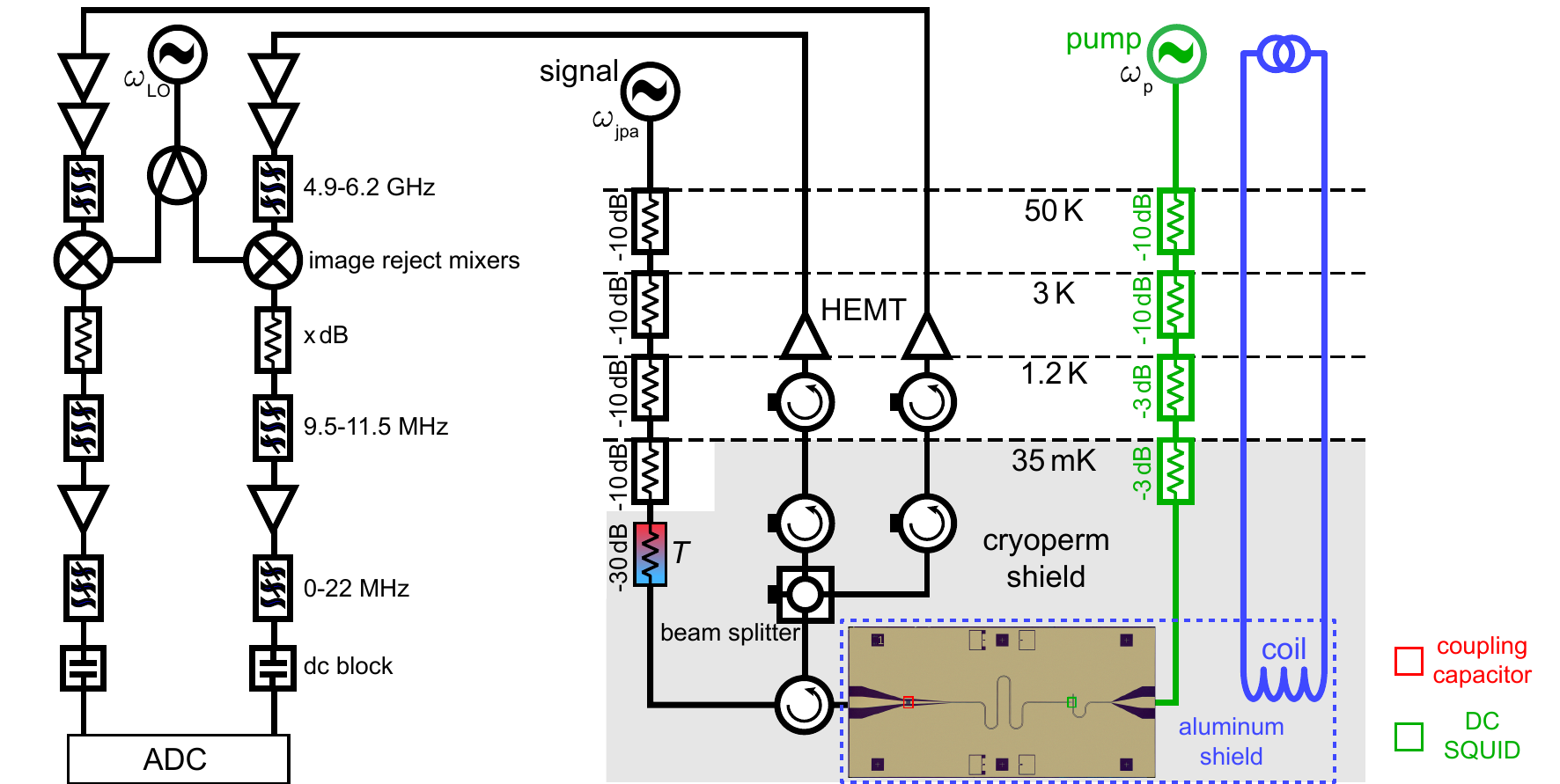}
\caption{\label{fig:S05} Schematics of the dual-path setup.}
\end{figure*}

\begin{figure*}[t]
\includegraphics{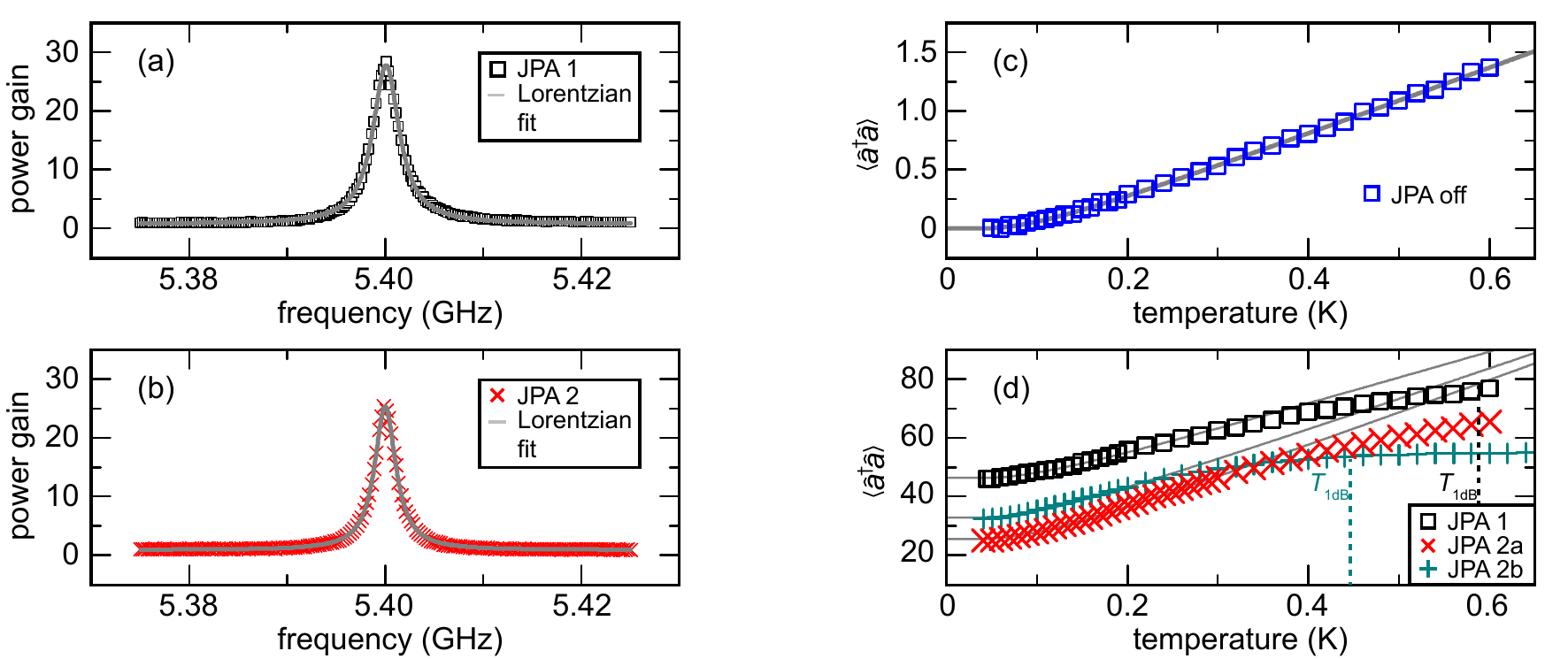}
\caption{\label{fig:S06} (a) JPA gain for a coherent drive measured in linear units versus frequency for JPA~1. The solid line is a Lorentzian fit used to determine the JPA bandwidth $B_{\mathrm{jpa}}\,{\simeq}\,\SI{3.2}{\mega\hertz}$. (b) As in panel (a) but for JPA~2 which has a bandwidth of \SI{2.6}{\mega\hertz}. (c) Photon number $\langle \hat{a}^{\dagger}\hat{a}\rangle$ at the input of the beam splitter measured in a Planck spectroscopy experiment versus temperature of the thermal emitter. The solid line is a fit of Eq.\,(\ref{eqn:n_nojpa}), which we use to determine the noise temperature and the gain of the amplification chain. (d) As in panel (c) but for the case that the JPA is turned on. By fitting Eq.\,(\ref{eqn:n_nojpa}) to the data points below \SI{200}{\milli\kelvin}, we obtain the gain and the noise photons added by the JPA. We also use this fit to determine the \SI{1}{\decibel} compression point indicated by the dashed line for JPA~2b.}
\end{figure*}

\paragraph{JPA sample} For the measurements based on the dual-path setup, we use the experimental setup presented in detail in the supplemental material of Ref.~\onlinecite{Fedorov_2016}. Both Josephson parametric amplifier (JPA) samples were designed and fabricated at NEC Smart Energy Research Laboratories, Japan and RIKEN, Japan. The dual-path setup comprises a flux-driven JPA with gain $G$ consisting of a quarter-wavelength transmission line resonator, which is short-circuited to ground by a DC SQUID (see Fig.\,\ref{fig:S05}). We couple an on-chip antenna inductively to the DC SQUID loop to apply a strong coherent pump tone $\omega_{\mathrm{p}}$ at approximately twice the resonant frequency $\omega_{\mathrm{jpa}}$ of the JPA. For the chosen working points the non-degenerate gains of the two JPAs have a bandwidth of approximately \SI{3}{\mega\hertz}, which we determine in a characterization measurement using a VNA [see Fig.\,\ref{fig:S06}\,(a) and (b)]. Additional specific parameters of the JPAs are summarized in Tab.\,\ref{tab:tab3}.\\
In our photon statistics experiments, we use a cryogenic hybrid ring as beam splitter to divide the signal into two amplification paths (dual-path method~\cite{Menzel_2010}). For uncorrelated input signals, i.e., vacuum and thermal states, the beam splitter does not affect the photon statistics of incident fields~\cite{Campos_1989}. After strong but independent amplification in the two paths, the signal is downconverted to an intermediate frequency $\omega_{\mathrm{if}}\,{=}\,\omega_{\mathrm{lo}}\,{-}\,\omega_{\mathrm{jpa}}\,{=}\,2\pi\,{\times}\,\SI{11}{\mega\hertz}$ and enters an analog-to-digital conversion (ADC) card. The particular digitizing procedure to calculate all correction moments $\langle I_{1}^{n}I_{2}^{m}Q_{1}^{k}Q_{2}^{\ell}\rangle$ up to fourth order ($0\,{\leq}\,n\,{+}\,m\,{+}\,k\,{+}\,\ell\,{\leq}\,4$ with $n,m,k,\ell\,{\in}\,\mathbb{N}_{0}$) is described in detail in Ref.~\onlinecite{Fedorov_2016}. From these calculations, we extract the signal moments $\langle(\hat{a}^{\dagger})^{n}\hat{a}^{m}\rangle$. Thermal states are generated as described in detail in the first section of this supplementary.\medskip

\begin{table*}[b]
\caption{\label{tab:tab3}Overview of the JPA samples. We perform one measurement with JPA\,1 and two individual measurements using JPA2 with different detuning $\delta_{\mathrm{jpa}}\,{=}\,\omega_{\mathrm{jpa}}\,{-}\,\omega_{\mathrm{p}}/2$ between JPA frequency $\omega_{\mathrm{jpa}}$ and pump frequency $\omega_{\mathrm{p}}$. The measurement bandwidth for all measurements is $\omega_{\mathrm{jpa}}\,{\pm}\,\SI{200}{\kilo\hertz}$. $({\star})$ Because the $\SI{1}{\decibel}$ compression point for JPA~1 is outside the measured temperature range, we can only estimate its value here.}
\begin{tabular}{cccccccccccccc}
\toprule
device & run & gain $G$ & $B_{\mathrm{jpa}}$ & $n_{\mathrm{n}}$ & $\rho$ & $\xi$ &  $\tilde{g}^{(2)}_{\mathrm{n}}(0)$ & $\delta_{\mathrm{jpa}}/2\pi$ & $\omega_{\mathrm{jpa}}/2\pi$ & $T_{1\mathrm{dB}}$ & $\mathcal{P}_{1\mathrm{dB}}$ & $\kappa_{\mathrm{x}}$ & $\kappa_{\mathrm{i}}$\\
JPA\,1 & -- & 14.3\,dB & \SI{3.2}{\mega\hertz} & 1.47 & 2.24 & 8.14 & 7.1 & \SI{100}{\kilo\hertz} & \SI{5.4}{\giga\hertz} & \SI{700}{\milli\kelvin}$^{\star}$  & -127\,dBm$^{\star}$  & \SI{18.7}{\mega\hertz} & \SI{5.4}{\mega\hertz}\\
JPA\,2 & a & 15.8\,dB & \SI{2.6}{\mega\hertz} & 0.66 & 2.23 & 3.29 & 1.1 & \SI{100}{\kilo\hertz} & \SI{5.4}{\giga\hertz} & \SI{590}{\milli\kelvin} & -129\,dBm & \SI{14.9}{\mega\hertz} & \SI{0.2}{\mega\hertz}\\
JPA\,2 & b & 15.2\,dB & \SI{2.6}{\mega\hertz} & 0.97 & 2.21 & 3.29 & 1.8 & \SI{500}{\kilo\hertz} & \SI{5.3}{\giga\hertz} & \SI{440}{\milli\kelvin} & -130\,dBm & \SI{14.6}{\mega\hertz} & \SI{0.2}{\mega\hertz}
\end{tabular}
\end{table*}

\paragraph{Photon number calibration} To calibrate the photon number $\langle \hat{a}^{\dagger}\hat{a}\rangle$ at the input of the hybrid ring, we first perform a Planck spectroscopy experiment with the JPA turned off to relate the detected power $\mathcal{P}_{\mathrm{det}}$ and photon number $\langle\hat{a}^{\dagger}\hat{a}\rangle\,{=}\,\exp\left(\hbar\omega/k_{\mathrm{B}}T\,{-}\,1\right)^{-1}$ via~\cite{Mariantoni_2010}
\begin{equation}
 \mathcal{P}_{\mathrm{det}} = G_{\mathrm{chain}}\times B \times \hbar\omega \left[\langle\hat{a}^{\dagger}\hat{a}\rangle + \frac{1}{2} +  k_{\mathrm{B}}T_{\mathrm{chain}}\right]\,.
\label{eqn:n_nojpa}
\end{equation}
Here, $G_{\mathrm{chain}}$ is the total amplification of the setup, $B\,{=}\,\SI{400}{\kilo\hertz}$ is the measurement bandwidth, and $T_{\mathrm{chain}}$ is the effective noise temperature of the amplification chain. As shown in Fig.\,\ref{fig:S06}\,(c), the data nicely follows Eq.\,(\ref{eqn:n_nojpa}) if the JPA is turned off. From this measurement, we obtain the noise temperature of the cryogenic amplifiers $T_{\mathrm{hemt}}\,{\approx}\,T_{\mathrm{chain}}\,{\simeq}\,\SI{3}{\kelvin}$ and the gain total $G_{\mathrm{chain}}\,{\simeq}\,\SI{145}{\decibel}$ of the amplification chain.
% When we turn on the JPAs, we record the power
% \begin{equation}
%  \mathcal{P}_{\mathrm{det}} = G_{\mathrm{chain}} \times B \times \hbar\omega\times \left\{G \left[\langle\hat{a}^{\dagger}\hat{a}\rangle + 1 + n_{\mathrm{n}}\right] +  k_{\mathrm{B}}T_{\mathrm{chain}}\right\}\,.
% \label{eqn:n_nojpa}
% \end{equation}
In order to characterize the JPA properties, we perform a temperature sweep when the JPA is turned on. As apperent from Fig.\,\ref{fig:S06}\,(d), there is a constant photon number offset due to the noise photons $n_{\mathrm{n}}$ added by the JPAs. Furthermore, the JPAs run into compression when the field temperature exceeds approximately $\SI{400}{\milli\kelvin}$. From the field temperature $T_{1\mathrm{dB}}$ at the \SI{1}{\decibel} compression point (cf.~Tab.\,\ref{tab:tab3}), we calculate \SI{1}{\decibel} values $\mathcal{P}_{1\mathrm{dB}}\,{=}\,\kappa_{\mathrm{x}}(2\pi)^{-1}k_{\mathrm{B}}T_{1\mathrm{dB}}\,{\simeq}\,\SI{-130}{\decibel}$m. Here, $\kappa_{\mathrm{x}}$ is the external coupling rate of the resonator, which strongly exceeds the internal loss rate $\kappa_{\mathrm{i}}$. The values obtained for $\mathcal{P}_{1\mathrm{dB}}$ fit well to the \SI{1}{\decibel} compression points measured for a coherent input state. For all measurements presented in this article, we use modest pump powers, such that we do not expect any non-linear effects~\cite{Kochetov_2015} of the JPAs.\medskip

\begin{figure*}[t]
\includegraphics{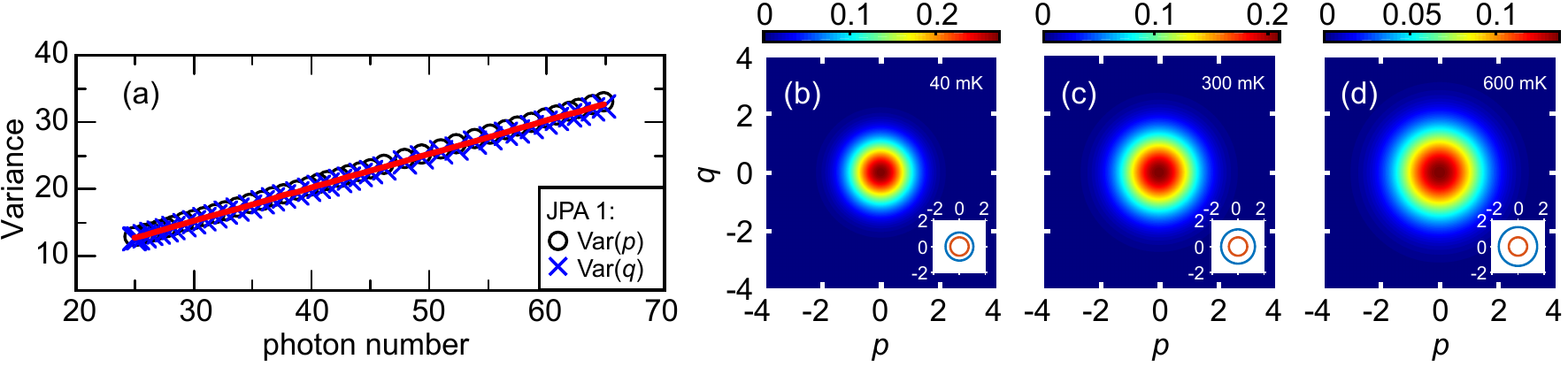}
\caption{\label{fig:S07} (a) Variance of $\hat{p}$ and of $\hat{q}$ plotted versus photon number. The solid line is a calculation based on Eq.\,(\ref{eqn:varpq}). (b)\,{-}\,(d) Wigner functions of thermal states measured with JPA on referenced back to the input of the hybrid ring. The insets show the $1/e$-contour of the vacuum (red) and the thermal state (blue).}
\end{figure*}

\paragraph{Variance of individual field quadratures} We use Eq.\,(\ref{eqn:Cjpa}) to describe the photon number variance of broadband amplified signals. When comparing the predicted values of $\xi\,{=}\,4n_{\mathrm{n}}\,{+}\,4$ and $\tilde{g}^{(2)}_{\mathrm{n}}(0)\,{=}\,2(n_{\mathrm{n}}\,{+}\,1)^{2}$, we observe that for both cases, the measured values are smaller than the expected values. To exclude that this effect is due to squeezing of the field quadratures, we analyze the variance of the individual quadrature components Var$(\hat{p})$ and Var$(\hat{q})$. Here, we define $\hat{p}\,{=}\,\imath(\hat{a}^{\dagger}\,{-}\,\hat{a})/2$ and $\hat{q}\,{=}\,(\hat{a}^{\dagger}\,{+}\,\hat{a})/2$. Then, at the input of the hybrid ring, one expects
\begin{equation}
 \frac{\mathrm{Var}(\hat{p})}{G} = \frac{\mathrm{Var}(\hat{q})}{G} = \frac{\langle\hat{a}^{\dagger}\hat{a}\rangle}{2} + \frac{1}{4}
 \label{eqn:varpq}
\end{equation}
for unsqueezed thermal states. As shown in Fig.\,\ref{fig:S07}\,(a), we observe the expected linear trend which fits very well to the expected behavior described in Eq.\,(\ref{eqn:varpq}). Hence, we do not observe any squeezing effects in the field quadratures of the amplified thermal fields. This circumstance is also expressed in the circular Wigner functions shown in Figs.\,\ref{fig:S07}\,(b)\,{-}\,(d).

% \bibliographystyle{apsrev-title-Frank}
% \bibliography{D:/Dropbox/Goetz_Bibliography}

\end{document}